   \documentclass[12pt]{iopart}
   \usepackage{iopams}
\newcommand{\Deqn}[1]{{(\ref{#1})}}
\newcommand{\Deqns}[1]{{(\ref{#1})}}

\newcommand{\beq}{\begin{equation}}
\newcommand{\eeq}{\end{equation}}

\newcommand{\bea}{\begin{eqnarray}}
\newcommand{\eea}{\end{eqnarray}}

\newcommand{\Text}[1]{{\mbox{#1}}}
\newcommand{\prim}[1]{{#1^\prime}}

\newcommand{\rd}{{\mbox{d}}}
\newcommand{\dx}{{\,\mbox{d}x}}
\newcommand{\dt}{{\,\mbox{d}t}}
\newcommand{\dr}{{\,\mbox{d}r}}
\newcommand{\rw}{{\mbox{rw}}}
\newcommand{\lz}{{\mbox{lz}}}

\newcommand{\HB}{\bar{H}}
\newcommand{\Lie}{{\pounds}}

\newcommand{\calB}{{\cal B}}
\newcommand{\calE}{{\cal E}}
\newcommand{\calI}{{\cal I}}
\newcommand{\calR}{{\cal R}}
\newcommand{\gotg}{\mathfrak{g}}

\newcommand{\lm}{{\ell m}}
\newcommand{\el}{{\ell}}

\newcommand{\sS}{{\mbox{\scriptsize S}}}
\newcommand{\R}{{\mbox{\scriptsize R}}}

\newcommand{\oo}{{\mbox{\scriptsize o}}}
\newcommand{\tail}{{\mbox{\scriptsize tail}}}
\newcommand{\dir}{{\mbox{\scriptsize dir}}}
\newcommand{\ret}{{\mbox{\scriptsize ret}}}
\newcommand{\adv}{{\mbox{\scriptsize adv}}}
\newcommand{\act}{{\mbox{\scriptsize act}}}
\newcommand{\rad}{{\mbox{\scriptsize rad}}}

\newcommand{\ev}{{\mbox{\scriptsize ev}}}
\newcommand{\od}{{\mbox{\scriptsize od}}}
\newcommand{\h}{{h}}
\newcommand{\trc}{{\mbox{\scriptsize trc}}}

\begin{document}
\title{Perspective on gravitational self-force analyses }
\author{  Steven Detweiler }
\address{ Department of Physics,\\ % PO Box 118440,\\
          University of Florida,\\
          Gainesville, FL 32611-8440}
\ead{det@ufl.edu}
%\date{10 March, 2005}

\begin{abstract}
A point particle of mass $\mu$ moving on a geodesic creates a perturbation
$h_{ab}$, of the spacetime metric $g_{ab}$, that diverges at the particle.
Simple expressions are given for the singular $\mu/r$ part of $h_{ab}$ and
its tidal distortion caused by the spacetime. This singular part $h^\sS_{ab}$
is described in different coordinate systems and in different gauges.
Subtracting $h^\sS_{ab}$ from $h_{ab}$ leaves a regular remainder
$h^\R_{ab}$. The self-force on the particle from its own gravitational field
adjusts the world line at $\Or(\mu)$ to be a geodesic of $g_{ab}+h^\R_{ab}$;
this adjustment includes all of the effects of radiation reaction.  For the
case that the particle is a small non-rotating black hole, we give a
uniformly valid approximation to a solution of the Einstein equations, with a
remainder of $\Or(\mu^2)$ as $\mu\rightarrow0$.

An example presents the actual steps involved in a self-force calculation.
Gauge freedom introduces ambiguity in perturbation analysis. However,
physically interesting problems avoid this ambiguity.
\end{abstract}


\section{Introduction}
 \label{introduction}

A description of motion always entails approximations and abstractions.
 The motion of a small black hole through spacetime is clearly not a geodesic
of the actual, physical spacetime geometry. After all, the ``center'' of a
black hole is inside the event horizon, where the geometry is unknown.
Nevertheless, if the mass of the hole is sufficiently small in comparison
with a length scale of spacetime, then the motion is approximately geodesic
on an abstract spacetime which is described as ``spacetime with the
gravitational field of the black hole removed''. Much of this manuscript
focuses upon the meaning of this last phrase.

In general relativity, an object of infinitesimal mass and size moves through
a background spacetime along a geodesic. If the particle has a small but
finite mass $\mu$ then its world line $\Gamma$ deviates from a geodesic of
the background by an amount proportional to $\mu$. This deviation is
sometimes described as resulting from the ``self-force'' of the particle's
own gravitational field acting upon itself and includes the effects which are
often referred to as radiation reaction.

In the literature the phrase ``gravitational self-force'' often refers to
precisely the right hand side of \Deqn{gravforce-a}, given below. As
emphasized by Barack and Ori \cite{BarackOri01}, the value of this quantity
depends upon the gauge being used (see section \ref{GaugeTransGSF}) and is,
thus, ambiguous. In this manuscript the phrase ``gravitational self-force''
is only used in an imprecise, generic way to describe any of the effects upon
an object's motion which are proportional to its own mass.

\subsection{Newtonian self-force example}
 \label{newtonian}
Newtonian gravity presents an elementary example of a self-force effect
\cite{DetPoisson04}. A small mass $\mu$ in a circular orbit of radius $R$
about a more massive companion $M$ has an angular frequency $\Omega$ given by
\beq
  \Omega^2 = \frac{M}{R^3(1+\mu/M)^2}.
\label{Kepler1}
\eeq
When $\mu$ is infinitesimal, the large mass $M$ does not move, the radius of
the orbit $R$ is equal to the separation between the masses and
$\Omega^2=M/R^3$. However when $\mu$ is finite but still small, both masses
orbit their common center of mass with a separation of $R(1+\mu/M)$, and the
angular frequency is as given in \Deqn{Kepler1}. The finite $\mu$ influences
the motion of $M$, which influences the gravitational field within which
$\mu$ moves. This back action of $\mu$ upon its own motion is the hallmark of
a self-force, and the $\mu$ dependence of \Deqn{Kepler1} is properly
described as a Newtonian self-force effect. When $\mu$ is much less than $M$,
an expansion of \Deqn{Kepler1} provides
\beq
  \Omega^2 % = \frac{\m}{r^3(1+\mu/M)^2}
       \approx \frac{M}{r^3}\left[1-2\mu/M + \Or(\mu^2/M^2)\right] .
\label{newtonianOm2}
\eeq
The finite mass ratio $\mu/M$ changes the orbital frequency by a fractional
amount
\beq
 \frac{\Delta\Omega}{\Omega} = -\frac{\mu}{M}.
\eeq
In this manuscript we describe any such $\Or(\mu/M)$ effect on the motion as
being a ``gravitational self-force'' effect. Below, we see that the
self-force effects for gravity include all of the consequences of what is
often referred to as ``radiation reaction.'' However, we also see that a
local observer, near $\mu$ deep inside the wave-zone and not privy to global
spacetime information, is unable to distinguish radiation reaction and the
other parts of the gravitational self-force from pure geodesic motion, at
this level of approximation.

\subsection{Electromagnetic radiation reaction in flat spacetime}

The Lorentz force law
\beq
  \mathbf{F} = q({\mathbf{E}} + {\mathbf{v}}\times{\mathbf{B}})
\label{Lorentz}
\eeq
describes the interaction of a point charge $q$ with an electromagnetic
field. In an elementary electricity and magnetism course, it is implicit that
$q$'s own electromagnetic field is not to be included on the right hand
side---after all for a point charge ${\mathbf{E}}$ is infinite at the very
location where it is to be evaluated in \Deqn{Lorentz}. Thus, the
electromagnetic field of \Deqn{Lorentz} is an ``external'' field, whose
source might be, say, the parallel plates of a capacitor but does not include
the charge $q$ itself.

Abraham and Lorentz first derived the radiation reaction force on a point
charge \cite{Jackson3rd}
\beq
 {\mathbf{F}}_\rad = \frac{2}{3} \frac{q^2}{c^3} \ddot{\mathbf{v}}
\label{ALDforce}
\eeq
in terms of the changing acceleration of $q$. This equation may be
interpreted in a perturbative sense: Let $q$ have a small mass and be
oscillating on the end of a spring. At lowest order in the perturbation, $q$
executes simple harmonic motion. At first order in the perturbation,  the
right hand side of \Deqn{ALDforce} is evaluated by use of a
$\ddot{\mathbf{v}}$ consistent with the harmonic motion. The resulting
${\mathbf{F}}$ is a small damping force which removes energy from the system
at just the proper rate to account for the outward energy flux of radiation.

A great value of \Deqn{ALDforce} resides in its elementary use by a theorist
to calculate the radiation reaction force.

A drawback of \Deqn{ALDforce} is the apparent obscuration of the root cause
of this force. Charges interact with electromagnetic fields via
\Deqn{Lorentz}. Yet, no electromagnetic field is present in \Deqn{ALDforce}.
Imagine a local observer extremely close to $q$, deep within the wave zone,
and with a length scale very much smaller than that associated with the
oscillations. This observer correctly interprets the majority of the
acceleration of $q$ as resulting from the coupling to the spring. The local
observer is unaware of the radiation---a non-local concept; yet, he must
explain the deviation from pure harmonic motion resulting from
${\mathbf{F}}_\rad$ as a consequence of the interaction of $q$ with some
external field via \Deqn{Lorentz}. The Abraham-Lorentz analysis correctly
\textit{calculates} the electromagnetic self-force. But it does not
\textit{explain} this self-force in terms of the charge interacting with an
external electromagnetic field.

Dirac \cite{Dirac38} removes this drawback by providing an interpretation of
\Deqn{ALDforce} as a direct consequence of \Deqn{Lorentz}, with the
electromagnetic field on the right hand side being an external field of
indeterminate origin to the local observer. Dirac uses the conservation of
the electromagnetic stress-energy tensor in a world-tube surrounding $q$, and
ultimately takes the limit of vanishing radius of the world-tube. One
consequence of his analysis is that the half-advanced plus half-retarded
field $F^\sS_{ab}=\frac12(F^\ret_{ab}+F^\adv_{ab})$ of $q$ exerts no force on
$q$ itself, even though the field is formally singular in the point charge
limit. We call the actual field $F^\act_{ab}$, and the remainder $F^\R_{ab} =
F^\act_{ab} - F^\sS_{ab}$ is a vacuum solution of Maxwell's equations.
$F^\R_{ab}$ substituted into the right hand side of \Deqn{Lorentz} yields
\Deqn{ALDforce}, as shown by Dirac.

A local observer measures the electromagnetic field in the vicinity of $q$,
but with no information regarding boundary conditions or distant radiation,
he can make no conclusions as to the detailed cause or source of the field.
However, in the perturbative sense described above, the observer can
calculate the singular field $F^\sS_{ab}$ in the vicinity of $q$. He can
subtract this singular field $F^\sS_{ab}$ from the actual, measured field
$F^\act_{ab}$ to obtain
\beq
  F^\R_{ab} =  F^\act_{ab} -  F^\sS_{ab} .
\eeq
The charge $q$ then interacts with the resulting regular source-free
electromagnetic field $F^\R_{ab}$ via \Deqn{Lorentz} with a resulting small
perturbation in its motion.  Thus, a local observer naturally explains the
damping of the harmonic motion as a consequence of $q$ interacting with an
external, locally source-free field $F^\R_{ab}$. However, with no global
information regarding boundary conditions he would not be able to determine
the source or cause of this external field. In particular the local observer
would see no phenomenon which he would be compelled to describe as radiation
reaction.

\subsection{Electromagnetic radiation reaction in curved spacetime}

DeWitt and Brehme's \cite{DeWittBrehme60} pioneering analysis of
electromagnetic radiation reaction in curved spacetime follows Dirac's
approach and also uses the conservation of energy in a world-tube to
determine the force on a point charge. Their results reduce to Dirac's in the
flat spacetime limit. However, DeWitt and Brehme find that generally
$\frac{1}{2}(F_{ab}^\ret+F_{ab}^\adv$) does, in fact, exert a force on the
charge in curved spacetime. After its removal from the actual field, the
remainder does not serve as the electromagnetic field on the right hand side
of \Deqn{Lorentz} for calculating a radiation reaction force.

To simplify the remainder of this introduction we, henceforth, assume that
the charge is in free fall in curved spacetime---the charge would move along
a geodesic except for interaction with its own electromagnetic field; there
are no springs attached.

DeWitt and Brehme use the Lorenz gauge, $\nabla_a A^a=0$, and a Hadamard
expansion to break the Green's function into the ``direct'' and ``tail''
parts with the vector potential
\begin{equation}
  A^\ret_{a} \equiv A^\dir_{a} + A^\tail_{a}.
\end{equation}
The direct part of the retarded Green's function has support only on the past
null cone, and the tail part has support only inside the past null cone. They
find that the electromagnetic self-force can be described as a consequence of
the particle interacting just with $A^\tail_{a}$,
\begin{equation}
  F_\rad^a = q g^{ac} (\nabla_c A^\tail_b - \nabla_b A^\tail_c ) u^b .
\label{emforce}
\end{equation}
This expression, like \Deqn{ALDforce}, has the great value that it can be
used to calculate an electromagnetic self-force, but it shares the drawback
that it does not explain the self-force in terms of a locally measurable,
source free solution of the Maxwell equations. In fact  $A^\tail_a$ is not in
any sense a solution of the electromagnetic field equation
\begin{equation}
 \nabla^2 A^a
           - {R^a}_b A^b
      = - 4\pi J^a .
\label{MaxEqn}
\end{equation}
The details of the Hadamard expansion reveal that if $A^a_\tail$ were
inserted into the left hand side here, it would yield a phantom $J_\tail^a$,
throughout a neighborhood of the charge. There would be no other evidence for
the existence of this $J_\tail^a$. Further, if $(R_{ab}-\frac{1}{6}g_{ab}R)
u^b \ne 0$, then $A^\tail_a$ is not differentiable at the particle and some
version of averaging around the charge is required to compute the self-force.
$A_\tail^a$ is a valuable mathematical construct which may be used to
calculate the self-force from \Deqn{emforce}, but it is not associated with
an actual electromagnetic field.  We conclude that the DeWitt-Brehme
construction correctly \textit{calculates} the electromagnetic self-force.
But it does not \textit{explain} the self-force in terms of the charge
interacting with an external electromagnetic field.

A modification \cite{DetWhiting03} of the DeWitt and Brehme analysis has
rectified this shortcoming. The actual vector potential may be decomposed as
\begin{equation}
  A^\act_{a} \equiv A^\sS_{a} + A^\R_{a},
\end{equation}
where $A^\sS_{a}$ and $A^\R_{a}$ are, in fact, solutions of Maxwell's
equations in a neighborhood of $q$: $A^\sS_{a}$ has only the charge $q$ as
its source, while $A^\R_{a}$ is a vacuum solution. Further, \Deqn{emforce}
yields the same force whether $A^\R_{a}$ or $A^\tail_{a}$ is inserted on the
right hand side, after the possible lack of differentiability of
$A^\tail_{a}$ is handled properly.

One nuance of the decomposition into S- and R-fields, is that the Green's
function for the S-field has support at the advanced and retarded times, just
as in the flat-spacetime example, above. But it also has support at the
events between the retarded and advanced times---these have a spacelike
separation with the field point.

The ``S'' and ``R'' decomposition provides a local observer in curved
spacetime with the ability to measure the actual electromagnetic field
$F^\act_{ab}$ in a neighborhood of $q$. He can make no conclusions as to the
detailed cause or source of the field.
 However, in the perturbative sense described above, the observer can
calculate $F^\sS_{ab}$ in a neighborhood of $q$ based upon its approximate
geodesic motion. He can then subtract this singular field $F^\sS_{ab}$ from
the actual, measured field $F^\act_{ab}$. The charge $q$ then interacts with
the resulting regular source-free electromagnetic field $F^\R_{ab}$ via
\Deqns{Lorentz} or (\ref{emforce}) with a resulting small perturbation of its
geodesic motion. Thus, a local observer naturally explains the lack of
geodesic motion of a charge $q$ as a consequence of $q$ interacting with an
external, locally source-free electromagnetic field. However, with no global
information regarding boundary conditions he is not able to determine the
source or cause of this external field. In particular, at this level of
approximation the local observer sees no phenomenon which he would be
compelled to describe as radiation reaction.

\subsection{Gravitational self-force}

The treatment of gravitational radiation reaction and self-force, in terms of
Green's functions, are formally very similar to that just described for the
electromagnetic field.

In some circumstances the gravitational field may be considered to have an
effective stress-energy tensor consisting of terms which are quadratic in the
derivatives of the metric.  Mino, Sasaki and Tanaka \cite{Mino97} follow the
DeWitt-Brehme \cite{DeWittBrehme60} approach, but with this gravitational
stress energy tensor. Ultimately, they conclude that the motion of a point
mass $\mu$ satisfies
\begin{equation}
  \mu u^b \nabla_b u^a = - \mu (g^{ab}+ u^a u^b) u^c u^d
     (\nabla_{c} h^\tail_{db} - \frac12 \nabla_b h^\tail_{cd}) .
 \label{gravforce-a}
\end{equation}
In an independent analysis within the same paper, they treat $\mu$ as a small
black hole moving in an external universe and use a general matched
asymptotic expansion to arrive at the same conclusion. In this latter
approach, the metric of the black hole is considered to be perturbed by the
external universe through which it is moving. Simultaneously, the metric of
the external universe is considered to be perturbed by the small mass $\mu$
moving through it. Others have used matched asymptotic expansions to describe
the motion of a small black in an external universe \cite{Manasse, Death,
Kates80, Alvi, det01, poisson:03}, but the connection between such results
and radiation reaction appears not to have been made before reference
\cite{Mino97}.

Quinn and Wald \cite{quinn-wald:97} use an axiom based analysis of the
gravitational self-force and also arrive at \Deqn{gravforce-a}.

The form of equation (\ref{gravforce-a}) is equivalent, through first order
in $h^\tail_{ab}$, to the geodesic equation for the metric
$g_{ab}+h^\tail_{ab}$. From one perspective then \Deqn{gravforce-a} is the
gravitational equivalent of \Deqn{emforce}. Equation (\ref{gravforce-a}),
like \Deqn{emforce}, has the great value that it can be used to calculate a
gravitational self-force, but it shares the drawback that it does not explain
the gravitational self-force in terms of geodesic motion in a locally
measurable, source free solution of the Einstein equations. In fact,
$h^\tail_{ab}$ is not in any sense a solution of the perturbed Einstein
equation, given below in \Deqn{EabTab}.

The details of the Hadamard expansion reveal that if $h^\tail_{ab}$ were
inserted into the left hand side of \Deqn{EabTab}, it would yield a phantom
stress-energy tensor $T^\tail_{ab}$, throughout a neighborhood of $\mu$.
There would be no other evidence for the existence of this $T^\tail_{ab}$.
Further, when $R_{acbd} u^c u^d \ne 0$, then $h^\tail_{ab}$ is not even
differentiable at the particle; although details reveal that averaging around
the particle is not required to compute the self-force with
\Deqn{gravforce-a}. $h^\tail_{ab}$ is a valuable mathematical construct which
may be used to calculate the self-force from \Deqn{gravforce-a}, but it is
not associated with an actual gravitational field.  We conclude that the
Mino, Sasaki and Tanaka and the Quinn and Wald constructions correctly
\textit{calculate} the gravitational self-force. But they do not
\textit{explain} the self-force in terms of geodesic motion in an external
gravitational field.

A modification \cite{DetWhiting03} of the analysis involving $h^\tail_{ab}$
has rectified this shortcoming. The actual metric perturbation may be
decomposed as
\begin{equation}
  h^\act_{ab} \equiv h^\sS_{ab} + h^\R_{ab},
\end{equation}
where $h^\sS_{ab}$ and $h^\R_{ab}$ are, in fact, solutions of the perturbed
Einstein equations \Deqn{EabTab} in a neighborhood of $\mu$: $h^\sS_{ab}$ has
only the mass $\mu$ as its source, while $h^\R_{ab}$ is a vacuum solution.
Further, \Deqn{gravforce-a} yields the same force whether $h^\R_{ab}$ or
$h^\tail_{ab}$ is inserted on the right hand side.

Earlier \cite{det01}, asymptotic matching was used to find an explicit
expression for the leading terms in an expansion of $h^\sS_{ab}$ in powers of
the distance away from $\mu$. Further, it was also shown that $h^\R_{ab} =
h^\act_{ab}-h^\sS_{ab}$ was at least $C^1$, with the given terms of the
expansion for $h^\sS_{ab}$, and that $\mu$ necessarily followed a geodesic of
$g_{ab}+h^\R_{ab}$ up to terms of $\Or(\mu^2/\calR^2)$, where $\calR$ is a
length scale of the background geometry. However, at that time it was
erroneously claimed \cite{det01} that the $h^\R_{ab}$ field was identical to
$h^\tail_{ab}$ because both led to the same equation of motion---namely
geodesic motion in $g_{ab}+h^\R_{ab}$. It was during a failing effort to
demonstrate directly this equivalence that the important differences between
the pair $h^\sS_{ab}$ and $h^\R_{ab}$ and the pair $h^\dir_{ab}$ and
$h^\tail_{ab}$ as possible solutions of the perturbed Einstein equations were
discovered \cite{DetWhiting03}.

A small mass $\mu$ moves through a background geometry $g_{ab}$ along a world
line $\Gamma$. At the lowest order in a perturbative sense, $\Gamma$ is a
geodesic. The Newtonian example given in section \ref{newtonian} implies that
$\Gamma$ deviates from geodesic motion in $g_{ab}$ by $\Or(\mu/\calR)$---it
is this deviation in which we are interested.

A local observer in curved spacetime has the ability to measure the actual
metric $g^\act_{ab}$ in a neighborhood of $\mu$. In a perturbative sense, the
observer can calculate $h^\sS_{ab}$ in a neighborhood of $\mu$ based upon its
approximately geodesic motion. He can then subtract this singular field
$h^\sS_{ab}$ from the actual, measured field $g^\act_{ab}$. The mass $\mu$
will be observed to move along a geodesic of
$g^\act_{ab}-h^\sS_{ab}=g_{ab}+h^\R_{ab}$. Thus, a local observer sees
geodesic motion of $\mu$ in the metric $g_{ab}+h^\R_{ab}$, which is a vacuum
solution of the Einstein equations, up to a remainder of $\Or(\mu^2)$ in a
neighborhood of $\mu$.
 With no global information regarding, say, the original background metric
$g_{ab}$, he would be unable to make any measurement which would distinguish
the separate parts $g_{ab}$ and $h^\R_{ab}$ which together make up the metric
through which $\mu$ is moving on a geodesic. At this level of approximation
the local observer sees only geodesic motion and no phenomenon which he would
be compelled to describe as radiation reaction.

\subsection{Outline}

Perturbation analysis, described in \ref{perturbation}, is the heart of the
self-force formalism.
 A variety of locally inertial coordinate systems are identified in
\ref{inertial}. Some of the ensuing mathematics is simplified by use of
notation, introduced in \ref{potential}, which is convenient for describing
vector and tensor harmonics in a spherically symmetric geometry.

Sections \ref{slow}-\ref{Spart} describe the metric in the neighborhood of a
small black hole as it moves through spacetime and provide an identification
of the singular ``S-part'' of a particle's gravitational field, which exerts
no force on the particle, itself.
 The remaining ``R-part'' of the particle's gravitational field is then seen
to be responsible for the gravitational self-force in \ref{Rpart}. The
confusion caused by the gauge freedom inherent in the R-part is summarized in
\ref{gauge}.

An example of a point mass in a circular orbit about a Schwarzschild black
hole reveals, in section \ref{example},  how the difficulty of gauge
dependence may be handled in carefully defined circumstances.
 Future prospects for gravitational self-force calculations are discussed in
\ref{conclusion}.

\subsection{Conventions and notation}
Conventions and notation are described here and again in context below. The
indices $a$, $b$, $c$\ldots are spacetime indices lowered and raised with the
metric $g_{ab}$ and its inverse; the derivative operator compatible with
$g_{ab}$ is $\nabla_a$. The metric of flat Minkowskii space is $\eta_{ab}$.
The indices $i$, $j$, $k$, $l$, $p$, $q$ are always used as spatial indices
and are raised and lowered with the flat three-metric $f_{ij}$. $\hat n_i$ is
a unit radial vector in flat space.

Indices $A$, $B, \dots$ are used to denote vector or tensor components which
are tangent to a two-sphere in spherically symmetric geometries, especially
those which are generated by ``potential'' functions as described in section
\ref{potential}. Spatial, symmetric trace-free tensors such as $\calE_{ij}$
or $\calB_{ijk}$ represent the external gravitational multipole moments, when
the gravitational field field is expanded in a locally inertial coordinate
system. The symbols $\calE$ and $\calB$ always refer to the even and odd
parity moments, respectively.   The scalars $\calE^{(2)}=\calE_{ij}\hat n^i
\hat n^j$ and $\calB^{(3)}=\calB_{ijk}\hat n^i \hat n^j \hat n^k$, for
examples, represent linear combinations of the $\el=2$ and $\el=3$ spherical
harmonics, respectively, which depend only upon the angles $\theta$ and
$\phi$ in the usual Schwarzschild coordinates, and are independent of $t$ and
$r$. the superscript $(2)$ denotes the value of $\el$.

A small particle of mass $\mu$ moves along a world line $\Gamma$
parameterized by the proper time $s$. $p$ is an event on $\Gamma$.
 $\calR$ is a representative length scale associated with a geodesic $\Gamma$
of spacetime---$\calR$ is the smallest of the radius of curvature, the scale
of inhomogeneities, and the time scale for changes in curvature along
$\Gamma$.
 We use $h^\sS_{ab}$to represent the singular source field, while $h^\mu_{ab}$
is an approximation to $h^\sS_{ab}$ based upon an asymptotic expansion.

\section{First order perturbation analysis}
 \label{perturbation}
Perturbation analysis provides the framework for an understanding of the
self-force and radiation reaction on an object of small mass and size in
general relativity. This begins with a background spacetime metric $g_{ab}$
which is a vacuum solution of the Einstein equations $G_{ab}(g)=0$. An object
of small mass $\mu$ then disturbs the geometry by an amount $h_{ab} =
\Or(\mu)$ which is governed by the perturbed Einstein equations with the
stress-energy tensor $T_{ab}=\Or(\mu)$ of the object being the source,
\begin{equation}
  E_{ab}(h) = - 8\pi T_{ab} +\Or(\mu^2).
\label{EabTab}
\end{equation}
Here $E_{ab}(h)$ is the linear, second order differential operator on
symmetric, two-indexed tensors schematically defined by
\begin{equation}
  E_{ab}(h) \equiv -\frac{\delta G_{ab}}{\delta g_{cd}} h_{cd},
\label{Eabdef}
\end{equation}
and $G_{ab}$ is the Einstein tensor of $g_{ab}$, so that
\begin{eqnarray}
  2E_{ab}(h) &=& \nabla^2 h_{ab} + \nabla_a \nabla_b h
           - 2 \nabla_{(a}\nabla^c h_{b)c}
\nonumber\\ & &
      + 2{R_a}^c{}_b{}^d h_{cd}  %% \nonumber\\&&
      + g_{ab} ( \nabla^c\nabla^d h_{cd} - \nabla^2 h ) ,
\label{Eab}
\end{eqnarray}
with $h \equiv h_{ab} g^{ab}$ and $\nabla_a$ and ${{R_a}^c}_b{}^d$ being the
derivative operator and Riemann tensor of $g_{ab}$.
 If $h_{ab}$ is a solution of \Deqn{EabTab}
then it follows from \Deqn{Eabdef} that $g_{ab}+h_{ab}$ is an approximate
solution of the Einstein equations with source $T_{ab}$,
\begin{equation}
  G_{ab}(g+h) = 8\pi T_{ab} + \Or(\mu^2).
\end{equation}

The Bianchi identity implies that
\beq
  \nabla^a E_{ab}(h) = 0
\eeq
for any symmetric tensor $h_{ab}$; this is discussed in \ref{bianchi}. Thus,
an integrability condition for \Deqn{EabTab} is that the stress-energy tensor
$T_{ab}$ be conserved in the background geometry $g_{ab}$,
\beq
  \nabla^a T_{ab} = \Or(\mu^2).
\label{DaTab}
\eeq

Perturbation analysis at the second order is no more difficult formally than
at the first. But the integrability condition for the second order equations
is that $T_{ab}$ be conserved not in the background geometry, but in the
first order perturbed geometry. Thus, before solving the second order
equations, it is necessary to change the stress-energy tensor in a way which
is dependent upon the first order metric perturbations. This modification to
$T_{ab}$ is said to result from the ``self-force'' on the object from its own
gravitational field and includes the dissipative effects of what is often
referred to as ``radiation reaction'' as well as other nonlinear aspects of
general relativity. This modification to $T_{ab}$ is $\Or(\mu^2)$ because
$T_{ab}$ itself is $\Or(\mu)$.

A description of general, $n$th order perturbation analysis is given in
 \ref{nthorder}. The procedure is similar to that just outlined. The
stress-energy tensor must be conserved with the metric $g^{(n-1)}_{ab}$ in
order to solve the $n$th order perturbed Einstein equation (\ref{Eabn}) for
$h^{(n)}_{ab}$. In an implementation, the task then alternates between
solving the equations of motion for the stress-energy tensor and solving the
perturbed Einstein equation for the metric perturbation. Similar alternation
of focus between the equations of motion and the field equations is present
in post-Newtonian analyses.

For many interesting situations the object is much smaller than the length
scale of the geometry through which it moves. We expect, then, that the
detailed structure of the source should be unimportant in determining its
subsequent motion.

To focus on those details of the self-force which are independent of the
object's structure we first attempt to model the object by an abstract point
particle with no spin angular momentum or internal structure.
 The stress-energy tensor of a point particle is
 \begin{equation}
  T^{ab} = \mu \int_{-\infty}^\infty \frac{ u^a u^b}{\sqrt{-g}}
       \delta^4(x^a-X^a(s)) \,\mbox{d}s
 \label{Tab}
 \end{equation}
where $X^a(s)$ describes the world line $\Gamma$ of the particle in some
coordinate system as a function of the proper time $s$ along the world line.

The naive replacement of a small object by a delta-function distribution for
the stress-energy tensor is satisfactory at first order in the perturbation
analysis. The integrability condition \Deqn{DaTab} requires the conservation
of the perturbing stress-energy tensor. For a point particle this implies
that the world line $\Gamma$ of the particle is an approximate geodesic of
the background metric $g_{ab}$, with $u^a\nabla_a u^b = \Or(\mu)$
(\textit{cf} \ref{geodesic}). The solution of \Deqn{EabTab} is formally
straightforward, even  for a distribution valued source. This procedure has
been used many times to study the emission of gravitational waves by a point
mass orbiting a black hole \cite{ReggeWheeler, DRPP, Zerilli}.

A difficulty appears with the second order integrability condition
(\ref{divTn}), with $n=2$. This condition seems to require that the particle
move along a geodesic of $g_{ab}+h_{ab}$. But $h_{ab}$ is singular precisely
at the location of the particle. To rectify this situation we look for a
method to identify and to remove the singular part $h^\sS_{ab}$ of the point
particle's metric perturbation and, thus, to find the remaining $h^\R_{ab}$.
We would then have the expectation that the point particle would move along a
geodesic of the abstract, perturbed geometry $g_{ab}+h^\R_{ab}$.

To avoid the singularity in $h_{ab}$, we replace the point particle
abstraction by a small Schwarzschild black hole. The difficulty caused by the
formal singularity is replaced by the requirement of boundary conditions at
the event horizon.
 Following Mino, Sasaki and Tanaka \cite{Mino97}, in section \ref{matching}
we use a matched asymptotic expansion to demonstrate how the $\Or(\mu)$
self-force adjusts the world line of the particle.
 For a small black hole moving in an external spacetime, the solution of the
Einstein equations divides into two overlapping parts: In the \textit{inner
region} near the black hole the metric is approximately the Schwarzschild
metric with a small perturbation caused by the external spacetime through
which it is moving. In the \textit{outer region} far from the black hole the
metric is approximately the background geometry of the external spacetime
with a small perturbation caused by the black hole. Let a length scale of the
background be $\calR$, and let $r$ be some measure of distance from the black
hole.
 Assume that $\mu \ll \calR$ so that the black hole is in a context
where it is meaningful to say that its mass is small.
 The inner region extends from the black hole out to $r \ll \calR$.
The outer region includes all $ r \gg \mu$. These two regions overlap in the
\textit{buffer region} where $\mu \ll r \ll \calR$.

When we focus on the inner region in sections \ref{slow} and \ref{Spart} the
object is a black hole, and we find an approximation for $h^\sS$ that
consists of the singular $\mu/r$ part of the Schwarzschild metric plus its
tidal distortion caused by the background geometry. Equations
(\ref{hS})-(\ref{h3muab}) give a straightforward approximation for
$h^\sS_{ab}$.
 When we focus on the
outer region we are free to think of the object as being a point particle.
Matching the perturbed metrics in the ``matching zone,'' within the buffer
region, in section \ref{matching} provides an approximate solution to the
Einstein equations with a remainder of $\Or(\mu^2/\calR^2)$, which is
uniformly valid in the limit $\mu/\calR\rightarrow 0$, everywhere outside the
event horizon as is demonstrated in section \ref{Rpart}.

The motion of the object is ultimately described as being geodesic in an
abstract metric $g_{ab}+h^\R_{ab}$, where $h^R_{ab}$ is the metric
perturbation which would result from a point particle, with the singular part
$h^\sS_{ab}$ removed. The majority of the remainder of this manuscript is the
elucidation of the steps which lead to the calculation of the $\Or(\mu)$
adjustment of a small object's world line.

\section{ Locally inertial coordinate systems}
\label{inertial}

A description of the metric perturbation $h_{ab}$ near a point mass $\mu$
moving along a geodesic $\Gamma$ is most convenient with coordinates in which
the background geometry looks as flat as possible at the location of the
particle. Let $\calR$ be a representative length scale of the background
geometry---the smallest of the radius of curvature, the scale of
inhomogeneities, and the time scale for changes in curvature along $\Gamma$.
Corresponding to any event $p$, there is always a \textit{locally inertial}
coordinate system for which the metric and the affine connection at $p$ are
those of flat Minkowskii space, $\eta_{ab}$. The value of the metric and its
first derivatives at $p$ in any coordinate system are all that is required to
determine a locally inertial system. The construction is described, for
example, by Weinberg \cite{Weinberg} in his equation (3.2.12). Locally
inertial coordinates at $p$ remain locally inertial under an inhomogeneous
Lorentz transformation. In addition, if $p$ is the origin of the coordinates,
then any transformation of the form $ x^a_\Text{\scriptsize new} =  x^a +
\lambda^a{}_{bcd} x^b x^c x^d$ is also a locally inertial system with the
origin at $p$. Such an $\Or( x^3)$ coordinate transformation changes the form
of the metric only by $\Or( x^2)$ in a neighborhood of $p$.

One specialization of locally inertial coordinates, which fixes the form of
the quadratic parts of the metric at $p$, are Riemann normal coordinates
\cite{MTW} where the metric takes the form
\beq
  g_{ab} = \eta_{ab} - \frac16 (R_{acbd}-R_{adbc}) x^c x^d
       + \Or(x^3/\calR^3).
\eeq
Any coordinate transformation of the form
\beq
   x^a_\Text{\scriptsize new} =  x^a
     +\lambda^a{}_{bcde} x^b x^c x^d x^e + \Or(x^5/\calR^4)
\eeq
preserves this Riemann normal form of the metric. The coordinate location of
an event $q$ is given in terms of a set of direction cosines, with respect to
orthonormal basis vectors at $p$, and the change in affine parameter along a
geodesic from $p$ to $q$. Riemann normal coordinates are defined only in a
region where the geodesics emanating from $p$ do not intersect elsewhere in
the region.

Coordinates $ x^a=(t, x, y, z)$ may be found which are locally inertial along
any geodesic $\Gamma$, with $t$ measuring the proper time $s$ on $\Gamma$. In
these coordinates $g_{ab} = \eta_{ab} + \Or( r^2/\calR^2)$, where $ r^2
\equiv  x^2 + y^2 + z^2 \equiv  x^i x_i$ and  the indices $i$, $j$, $k$, $l$,
$p$, $q$ run over the spatial coordinates $x$, $y$ and $z$. A coordinate
transformation of the form $ x^a_\Text{\scriptsize new} =  x^a +
\lambda^a{}_{ijk}(s) x^i  x^j  x^k + \Or( r^4/\calR^3)$ preserves these
features with most components of the metric changing by $\Or( r^2/\calR^2)$.
However, $g_{tt}$ changes only by $\Or( r^3/\calR^3)$ and is always of the
simple form
 $g_{tt} = -1 - R_{titj} x^i  x^j + \Or( r^3/\calR^3)$, where
$R_{titj}$ is evaluated on $\Gamma$.

\subsection{Fermi normal coordinates}
Fermi normal coordinates \cite{ManasseMisner} are one specialization of
locally inertial coordinates on a geodesic $\Gamma$ for which the $\Or(
r^2/\calR^2)$ parts of the metric have a particularly appealing form as
simple combinations of components of the Riemann tensor evaluated on
$\Gamma$, \cite{MTW}
\bea
  g_{ab} \dx^a \dx^b &=& -(1 + R_{titj} x^i  x^j) \dt^2
% \nonumber\\ && {}
      - \frac{4}{3} R_{tikj} x^i  x^j \dt \, \dx^k
\nonumber\\ && {}
      + (f_{kl} - \frac{1}{3} R_{kilj} x^i  x^j) \dx^k \dx^l
\nonumber\\ && {}
      + \Or( r^3/\calR^3).
\label{FNmetric}
\eea
Li and Ni \cite{LiNi79} give the form of the metric in Fermi normal
coordinates to higher order.  The defining characteristics of Fermi normal
coordinates are that they are orthogonal on $\Gamma$, that the spatial axes
are geodesics, and that the distance from $\Gamma$ at proper time $s$ to an
event $(t=s,x^i)$ is $(x^i x^j \delta_{ij})^{1/2}$, when measured along a
geodesic perpendicular to $\Gamma$.

\subsection{THZ Normal coordinates}
 \label{THZcoords}
A second specialization of locally inertial coordinates on $\Gamma$,
introduced by Thorne and Hartle \cite{ThorneHartle85} and extended by Zhang
\cite{Zhang86}, describe the external multipole moments, defined on $\Gamma$,
of a vacuum solution of the Einstein equations. In these \textit{THZ
coordinates}
\begin{eqnarray}
  g_{ab} &=& \eta_{ab} + H_{ab}
\nonumber\\
     &=&  \eta_{ab} + {}_2H_{ab} + {}_3H_{ab} + \Or( r^4/\calR^4),
\label{Hs}
\end{eqnarray}
with
\begin{eqnarray}
\fl  {}_2H_{ab}d x^a d x^b & = &
         - \calE_{ij}  x^i  x^j ( \dt^2 + f_{kl} \rd x^k \dx^l )
% \nonumber\\ &&
         {} + \frac{4}{3} \epsilon_{kpq}\calB^q{}_i  x^p  x^i \dt \dx^k
\nonumber\\ \fl &&
         {} - \frac{20}{21} \Big[ \dot{\calE}_{ij}x^ix^j x_k
            - \frac{2}{5}  r^2 \dot{\calE}_{ik} x^i \Big] \dt \dx^k
\nonumber\\ \fl &&
 {} + \frac{5}{21} \Big[ x_i \epsilon_{jpq} \dot{\calB}^q{}_k x^px^k
    -  \frac{1}{5}  r^2 \epsilon_{pqi}
                 \dot{\calB}_{j}{}^q x^p \Big] \,\dx^i\, \dx^j
    +  \Or( r^4/\calR^4)
\label{H2}
\end{eqnarray}
and
\begin{eqnarray}
  {}_3H_{ab}\rd x^a \rd x^b & = &
         - \frac13\calE_{ijk} x^i x^j x^k
            ( \dt^2 + f_{kl} \dx^k \dx^l )
\nonumber\\ &&
         {} + \frac{2}{3} \epsilon_{kpq}\calB^q{}_{ij}
             x^p x^i x^j \dt \dx^k
          +  \Or( r^4/\calR^4),
\label{H3}
\end{eqnarray}
where $\epsilon_{ijk}$ is the flat space Levi-Civita tensor. These
coordinates are well defined up to the addition of arbitrary functions of
$\Or( r^5/\calR^4)$. The \textit{external multipole moments} $\calE_{ij}$,
$\calB_{ij}$, $\calE_{ijk}$, and $\calB_{ijk}$ are spatial, symmetric,
tracefree (STF) tensors and are related to the Riemann tensor evaluated on
$\Gamma$ by
\beq
 \calE_{ij} = R_{titj},
\eeq
\beq
  \calB_{ij} = \epsilon_i{}^{pq}R_{pqjt}/2,
\eeq
\beq
  \calE_{ijk} =\left[\partial_kR_{titj}\right]^{\mbox{\scriptsize STF}}
\eeq
and
\beq
   \calB_{ijk} = \frac{3}{8}
     \left[\epsilon_i{}^{pq}\partial_kR_{pqjt}
                  \right]^{\mbox{\scriptsize STF}},
\eeq
where ${}^{\mbox{\scriptsize STF}}$ means to take the symmetric, tracefree
part with respect to the spatial indices. $\calE_{ij}$ and $\calB_{ij}$ are
$\Or(1/\calR^2)$, while $\calE_{ijk}$ and $\calB_{ijk}$ are $\Or(1/\calR^3)$.
The dot denotes differentiation of the multipole moment with respect to $t$
along $\Gamma$. Thus $\dot\calE_{ij}=\Or(1/\calR^3)$ because $\calR$ limits
the time scale along $\Gamma$. All of the above external multipole moments
are tracefree because the background geometry is assumed to be a vacuum
solution of the Einstein equations.

The THZ coordinates are a specialization of harmonic coordinates, and
it is useful to define the ``Gothic'' form of the metric
\beq
  \gotg^{ab} \equiv \sqrt{-g} g^{ab}
\label{gotgdef}
\eeq
as well as
\beq
  \HB^{ab} \equiv \eta^{ab} - \gotg^{ab} .
\label{HBdef}
\eeq
A coordinate system is harmonic if and only if
\beq
  \partial_a \HB^{ab} = 0.
\label{HBdiv}
\eeq
Zhang \cite{Zhang86} gives an expansion of $\gotg^{ab}$ for an arbitrary
solution of the vacuum Einstein equations in THZ coordinates, his
equation~(3.26). The terms of $\HB^{ab}$ in this expansion include
\beq
 \HB^{ab} = {}_2\HB^{ab} + {}_3\HB^{ab} + \Or( r^4/\calR^4)
\label{HBab}
\eeq
where
\begin{eqnarray}
  {}_2\HB^{tt}& = & - 2 \calE_{ij}  x^i  x^j
\nonumber\\
  {}_2\HB^{tk}& = & - \frac{2}{3} \epsilon^{kpq}\calB_{qi} x_p  x^i
         {} + \frac{10}{21} \Big[ \dot{\calE}_{ij}x^ix^j x^k
            - \frac{2}{5} \dot{\calE}_{i}{}^k x^i  r^2\Big]
\nonumber\\
  {}_2\HB^{ij}& = & \frac{5}{21} \Big[ x^{(i}
       \epsilon^{j)pq} \dot{\calB}_{qk} x_p x^k
    -  \frac{1}{5} \epsilon^{pq(i} \dot{\calB}^{j)}{}_q x_p  r^2 \Big]
\label{HB2}
\end{eqnarray}
and
\begin{eqnarray}
  {}_3\HB^{tt} & = &
         - \frac23\calE_{ijk}  x^i  x^j  x^k
\nonumber\\
  {}_3\HB^{tk} & = & - \frac{1}{3} \epsilon^{kpq}\calB_{qij} x_p  x^i x^j
\nonumber\\
  {}_3\HB^{ij} & = & \Or( r^4/\calR^4) .
\label{HB3}
\end{eqnarray}

If $ r/\calR \ll 1$ then $H_{ab}$ is approximately the trace reversed version
of $\HB^{ab}$,
\beq
H_{ab} = \HB_{ab} - \frac12 \eta_{ab} \HB^c{}_c + \Or( r^4/\calR^4),
\eeq
and \Deqns{Hs}-(\ref{H3}) correspond precisely to (\ref{HBab})-(\ref{HB3}) up
to a remainder of $\Or( r^4/\calR^4)$.

Zhang \cite{Zhang86} gives the transformation from Fermi normal coordinates
to the THZ coordinates
\bea
  t_\Text{\scriptsize thz} &=& t_\Text{\scriptsize fn}
\nonumber\\
   x^i_\Text{\scriptsize thz} &=&   x^i_\Text{\scriptsize fn}
      - \frac{ r^2}{6} \calE^i{}_j  x^i_\Text{\scriptsize fn}
   + \frac{1}{3} \calE_{jk}  x^j_\Text{\scriptsize fn}
          x^k_\Text{\scriptsize fn}  x^i_\Text{\scriptsize fn}
          + \Or(r^4/\calR^3).
\label{thzfn}
\eea

\subsection{An application of THZ coordinates}

The scalar wave equation takes a particularly simple form in THZ coordinates,
\bea
  \sqrt{-g}\nabla^a\nabla_a\psi &=&
     \partial_a \big( \sqrt{-g}g^{ab} \partial_b \psi \big)
\nonumber\\  &=& {}
         \partial_a \big( \eta^{ab} \partial_b \psi \big)
       - \partial_a \big( \HB^{ab}  \partial_b \psi \big)
\nonumber\\  &=& {}
         (\eta^{ab} - \HB^{ab}) \partial_a \partial_b \psi,
\eea
where the second equality follows from \Deqn{HBdef} and the last from
\Deqn{HBdiv}. After an expansion of the contractions on $\HB^{ab}$, this
becomes
\beq
  \sqrt{-g}\nabla^a\nabla_a\psi =
         \eta^{ab} \partial_a \partial_b \psi
       -  \HB^{ij} \partial_i \partial_j \psi
       - 2 \HB^{it} \partial_{(i}  \partial_{t)} \psi
       -  \HB^{tt} \partial_t \partial_t \psi .
\label{del2psiHB}
\eeq
An approximate solution $\psi$ with a point charge source is $q/ r$. Direct
substitution into \Deqn{del2psiHB} reveals just how good this approximation
is. If $\psi$ is replaced by $q/ r$ on the right hand side, then the first
term gives a $\delta$-function, the third and fourth terms vanish because $
r$ is independent of $t$, and in the second term ${}_2\HB^{ij}$ has no
contribution because of the details given in \Deqn{HB2}, and the $\Or(
r^4/\calR^4)$ remainder of $\HB^{ij}$ yields a term that scales as $\Or(
r/\calR^4)$. Thus,
\beq
  \sqrt{-g}\nabla^a\nabla_a(q/ r) = -4\pi q\delta^3(x^i)
          +  \Or( r/\calR^4) .
\label{del2rho}
\eeq
Note that the remainder $\Or( r/\calR^4)$ is $C^0$. From the consideration of
solutions of Laplace's equation in flat spacetime,
% in  \ref{hRisC1}
it follows that a $C^2$ correction to $q/ r$, of $\Or( r^3/\calR^4)$, would
remove the $\Or( r/\calR^4)$ remainder on the right hand side. We conclude
that $q/ r + \Or( r^3/\calR^4)$ is a solution of the scalar field wave
equation for a point charge and that the error in the approximation of the
solution by $q/ r$ is $C^2$. In Ref.~\cite{DetMessWhiting03} we show that $q/
r$ is the singular field $\psi^\sS$ for a scalar charge, up to a remainder of
$\Or( r^3/\calR^4)$. This was done by use of a Hadamard expansion of the
Green's function.

THZ coordinates provide elementary, approximate solutions to the wave
equation with a singular source for vector and tensor fields as well
\cite{Messaritaki03} .

\section{Vector and tensor harmonics}
 \label{potential}

The forms of ${}_2H_{ab}$ and ${}_3H_{ab}$ in \Deqns{H2} and (\ref{H3}) might
appear unfamiliar, but they actually consist of $\el=2$ and 3 vector and
tensor spherical harmonics and have a close relationship with those
introduced by Regge and Wheeler \cite{ReggeWheeler} in their analysis of
metric perturbations of Schwarzschild black holes.
 This relationship is clarified with an example of $\calE_{ij}$, whose
Cartesian components are symmetric, tracefree, and constant. However, the
spherical-coordinate component $\calE_{rr}$ has the angular dependence of a
linear combination of the $Y_\lm$'s for $\el=2$. Thus, it is convenient to
define $\calE^{(2)} \equiv \calE_{ij}\hat n^i \hat n^j$, where $\hat n^i$ is
the unit radial vector in flat space. $\calE^{(2)}$ is a scalar field which
carries all of the information contained in the constant Cartesian components
of $\calE_{ij}$ and may be used to generate related quadrupole vector and
tensor harmonics.

For the angular components of vectors and tensors, we find it convenient to
follow Thorne's description of the pure-spin vector and tensor harmonics
\cite{ThorneRMP}, which are closely related to the harmonic decomposition
used by Regge and Wheeler \cite{ReggeWheeler}. For example, the spin-1 vector
harmonics generated by the spherical harmonic function $Y_\lm$ are the even
parity
\beq
   Y^{\mbox{\scriptsize E}\lm}_a =  r \sigma_a{}^b \nabla_b Y_\lm
\eeq
and the odd parity
\beq
   Y^{\mbox{\scriptsize B}\lm}_a
          = - r \epsilon_a{}^b \nabla_b Y_\lm,
\eeq
where
\beq
  \sigma_{ab} \equiv g_{ab} + u_a u_b - n_a n_b
\eeq
is the metric of a constant $t$,$r$ two-sphere, and
\beq
  \epsilon_{ab} \equiv \epsilon_{trab},
  \quad \mbox{with} \quad
  \epsilon_{tr\theta\phi} = \epsilon_{\theta\phi} = r^2\sin\theta,
\eeq
is the Levi-Civita tensor on the same two-sphere. Here $u_a$ and $n_a$ are
the unit normals of surfaces of constant $t$ and constant $r$, respectively.

We generalize this approach: For a vector field $\xi_a$, the parts
$\sigma_a{}^b \xi_b$ which are tangent to a two-sphere may be described by
two ``potentials'' $\xi^\ev$ and $\xi^\od$ via
\beq
   \sigma_a{}^b \xi_b = r \sigma_a{}^b \nabla_b \xi^\ev
                                   - r \epsilon_a{}^b \nabla_b \xi^\od .
\label{perpxi}
\eeq
The potentials $\xi^\ev$ and $\xi^\od$ are generally functions of all of the
spacetime coordinates and are guaranteed to exist by the invertibility of the
two dimensional Laplacian on a two-sphere. The factors of $r$ are included
for convenience.

The notation for a covariant vector field is condensed by defining even and
odd parity vectors associated with the potential $\xi^\ev$
\beq
 \xi^\ev_a \equiv  r \sigma_a{}^b \nabla_b \xi^\ev
\label{xiev}
\eeq
and with the potential $\xi^\od$
\beq
 \xi^\od_a \equiv - r \epsilon_a{}^b \nabla_b \xi^\od .
\label{xiod}
\eeq
The four independent components of a covariant vector in a spherically
symmetric geometry may be written as a sum of the form
\beq
  \xi_a \dx^a = \xi_t \dt + \xi_r \dr
                + \left(\xi^\ev_A + \xi^\od_A\right) \dx^A
\eeq
in terms of the four functions $\xi_t$, $\xi_r$, $\xi^\ev$ and $\xi^\od$. The
capital index $A$ is used here just as a reminder that the vector to which it
is attached is tangent to the two-sphere. The $A$ index should otherwise be
considered an ordinary spacetime index in the covariant spirit of
\Deqn{perpxi}-\Deqn{xiod}.

Similarly for a symmetric tensor field $h_{ab}$, the parts which are tangent
to a two-sphere $\sigma_a{}^c \sigma_b{}^d h_{cd}$ may be described by the
trace with respect to $\sigma_{ab}$ and by two potentials $h^\ev$ and $h^\od$
via
\bea
   \sigma_a{}^c \sigma_b{}^d h_{cd}
      &=& \frac{1}{2} \h^\trc \sigma_{ab}
    + r^2 \sigma_{(a}{}^c \sigma_{b)}{}^d \nabla_c
                      \left( \sigma_d{}^e \nabla_e  \h^\ev\right)
    - \frac{1}{2} r^2 \sigma_{ab} \sigma^{cd} \nabla_c
        \left(\sigma_d{}^e \nabla_e  \h^\ev\right)
\nonumber\\ &&
   {} -  r^2 \epsilon_{(a}{}^{c} \sigma_{b)}{}^{d} \nabla_c
       \left(\sigma_d{}^e \nabla_e \h^\od\right)
\label{perph}
\eea
The potentials $h^\ev$ and $h^\od$ are generally functions of all of the
spacetime coordinates and are guaranteed to exist by theorems involving
solutions of elliptic equations on a two-sphere. The factors of $r^2$ are
included for convenience.

The notation for a covariant tensor field is condensed by defining trace-free
tensors tangent to a two-sphere and associated with the potential $h^\ev$
\beq
 \h^\ev_{ab} \equiv
    r^2 \sigma_{(a}{}^c \sigma_{b)}{}^d \nabla_c
              \left( \sigma_d{}^e \nabla_e  \h^\ev\right)
    - \frac{1}{2}r^2 \sigma_{ab} \sigma^{cd} \nabla_c
        \left(\sigma_d{}^e \nabla_e  \h^\ev\right)
\label{hevAB}
\eeq
and with the potential $h^\od$
\beq
  \h^\od_{ab} \equiv - r^2 \epsilon_{(a}{}^{c} \sigma_{b)}{}^{d} \nabla_c
       \left(\sigma_d{}^e \nabla_e \h^\od \right) .
\label{hodAB}
\eeq
The ten independent components of a symmetric covariant tensor $h_{ab}$ in a
spherically symmetric geometry may be written as a sum of the form
\bea
   h_{ab} \dx^a \dx^b &=&
        \h_{tt} \dt^2 + 2 \h_{tr} \dt \dr
          + 2 \left( \h^\ev_{tA} + \h^\od_{tA}\right) \dt \dx^A
\nonumber\\ &&
  {} + \h_{rr} \dr^2 + 2 \left(\h^\ev_{rA} +  \h^\od_{rA}\right)  \dr \dx^A
\nonumber\\ &&
   {}  + \left( \frac{1}{2} \h^\trc \sigma_{AB}
          +  \h^\ev_{AB} +  \h^\od_{AB}\right) \dx^A \dx^B
\eea
in terms of the ten functions $h_{tt}$, $h_{tr}$, $h^\ev_{t}$, $h^\od_{t}$,
$h_{rr}$, $h^\ev_{r}$, $h^\od_{r}$, $h^\trc$, $h^\ev$ and  $h^\od$,
   As with the vector field, the capital indices $A$ and
$B$ are used here just as a reminder that the vector or tensor to which they
are attached is tangent to the two-sphere. Otherwise, they should be
considered ordinary spacetime indices in the covariant spirit of
\Deqn{perph}-\Deqn{hodAB}.

The descriptions of vector and tensor potentials in \Deqns{perpxi} and
(\ref{perph}) on a two-sphere could have been written with a derivative
operator involving the usual angular coordinates. However, this would cloud
the covariant nature of the decomposition which is clearly revealed above.

The description of the vector and tensor components in terms of potentials
takes advantage of the natural symmetry of the background geometry. For
example, if a potential is a function of $r$ and $t$ times a $Y_\lm$ then the
resulting vector or tensor field is the same function times the vector or
tensor spherical harmonic with the same $\el,m$ pair. Expressions such as the
perturbed Einstein tensor take a particularly simple form when written in
terms of the potentials in place of the components.

We assume throughout that $\calE$ is always associated with even parity
vectors and tensors, and that $\calB$ is always associated with odd parity
vectors and tensors. Thus, ${}^\ev$ and ${}^\od$ are often understood in
$\calE=\calE^\ev$ or $\calB=\calB^\od$.
 A superscript in parentheses, as in $\calE^{(2)}=\calE_{ij} n^{ij}$, denotes
the multipole index $\el$ which is also the number of indices in the STF
tensor $\calE_{ij}$.

With this notation, alternative forms of \Deqn{H2} and (\ref{H3}) are
\bea
  {}_2H_{ab}\dx^a \dx^b &=&  - r^2\calE^{(2)}
   \left( \dt^2 + \dr^2 + \sigma_{AB}\dx^A\dx^b \right)
       + 2\frac{r^2}{3} \calB^{(2)}_{A} \dt\dx^A
\nonumber\\  &&{}
 - 2\frac{2r^3}{7} \dot\calE^{(2)}   \dt \dr
 + 2\frac{2r^3}{21} \dot\calE^{(2)}_{A}  \dt \dx^A
\nonumber\\  &&{}
    + 2 \frac{r^3}{21} \dot\calB^{(2)}_{A}  \dr \dx^A
    - \frac{r^3}{42} \dot\calB^{(2)}_{AB}  \dx^A \dx^B
          +  \Or(r^4/\calR^4)
\label{H2new}
\eea
and
\begin{eqnarray}
  {}_3H_{ab}\dx^a \dx^b & = &
         - \frac{r^3}{3}\calE^{(3)}
             (\dt^2 + \dr^2 + \sigma_{AB}\dx^A\dx^b)
\nonumber\\ &&
         {} + 2\frac{r^3}{9} \calB^{(3)}_A \dt \dx^A
          +  \Or(r^4/\calR^4) .
\label{H3new}
\end{eqnarray}

\section{Slowly time dependent perturbations of the Schwarzschild geometry}
 \label{slow}

When a small Schwarzschild black hole of mass $\mu$ moves through a
background spacetime, the hole's metric is perturbed by tidal forces arising
from $H_{ab}$ in \Deqn{Hs}, and the actual metric near the black hole is
\begin{equation}
  g^\act_{ab} = g_{ab}^{\Text{\scriptsize Schw}} + {}_2h_{ab} + {}_3h_{ab}
   + \Or( r^4/\calR^4),
\label{gabpert}
\end{equation}
where the quadrupole metric perturbation ${}_2h_{ab}$ is a solution of the
perturbed Einstein equations \Deqn{EabTab}. The appropriate boundary
conditions for ${}_2h_{ab}$ are that it's components be well behaved on the
future event-horizon, in a well-behaved coordinate system, and that
${}_2h_{ab} \rightarrow {}_2H_{ab}$ in the {\it buffer region}
\cite{ThorneHartle85}, where $\mu \ll r \ll \calR$. The octupole metric
perturbation ${}_3h_{ab}$ has a similar description.

In \ref{SloMo} we follow Poisson's recent analysis
\cite{Poisson04a,Poisson04b,Poisson05} of a tidally distorted black hole, and
describe the metric perturbation for $r \ll \calR$ in
\Deqn{Aapp}-\Deqn{Japp}. An expansion of the metric perturbation in the
buffer region for
 $\mu \ll r \ll \calR$ ultimately provides the even parity
\bea \fl
  {}_2h^\ev_{ab} \dx^a \dx^b &=& - \calE^{(2)}
      \left[(r - 2\mu)^2 \dt^2 +r^2 \dr^2 + (r^2-2\mu^2)\sigma_{AB} \dx^A
      \dx^B\right]
\nonumber\\ \fl &&
  {} + \frac{16\mu^6}{15r^4} \dot\calE^{(2)}
      \left[  2(r+\mu) \dt^2
            + 2(r+5\mu) \dr^2
            + (2r+5\mu)\sigma_{AB} \dx^A\dx^B\right]
\nonumber\\ \fl && % h_{tr} =
 {} - 2 \frac{r(2r^3 - 3\mu r^2 - 6\mu^2r + 6\mu^3)}{3(r-2\mu)}
                                     \dot\calE^{(2)} \dt\dr
  +  \Or(\mu^8\dot\calE^{(2)}/r^5),
\label{h2ev}
\eea
and the odd parity
\bea
\fl {}_2h^\od_{ab} \dx^a \dx^b &=&
       2 \left[ \frac{r}{3}(r-2\mu) \calB^{(2)}_A
              + \frac{16\mu^6}{45r^4} (3r+4\mu) \dot\calB^{(2)}_A \right]
     \dt \dx^A
\nonumber\\ \fl &&  %  h^\od_{rA} =
 {} + 2 \frac{r^4}{12(r - 2\mu)} \dot\calB^{(2)}_A \dr \dx^A
  + \Or(\mu^8 \dot\calB^{(2)}/r^5),
\label{h2od}
\eea
which together properly match the $\Or(r^2/\calR^2)$ terms of (\ref{H2}) or
of (\ref{H2new}); the $\Or(r^3/\calR^3)$ terms are in a different gauge.
 In this form $\calE^{(2)}$ and $\calB^{(2)}$ are considered functions of $t$
and $\dot\calE^{(2)}$ denotes the $t$ derivative of $\calE^{(2)}$. Together,
these provide the quadrupole metric perturbation up to remainders of
$\Or(r^4/\calR^{4})$ and $\Or(\mu^8/r^5\calR^{3})$.

The approximately time independent octupole perturbation ${}_3H_{ab}$ of the
small black hole may be treated similarly. The time independent solution of
$E^{\Text{\scriptsize Schw}}_{ab}({}_3h) = 0$ which is well behaved on the
event horizon and properly matches the $\Or(r^3/\calR^3)$ terms of \Deqn{H3}
or of (\ref{H3new})
\begin{eqnarray}
  &&\lefteqn{ {}_3h_{ab}dx^a \dx^b  =
         - \frac{r^3}{3}\calE^{(3)} \left[
                \left(1-\frac{2\mu}{r}\right)^2\left(1-\frac{\mu}{r}\right)
 \dt^2 \right.
                }
\nonumber  \\ & & \left.
       \quad  {}+ \left(1-\frac{\mu}{r}\right) \dr^2
         + \left(r^2-2\mu r + \frac{4\mu^3}{5r}\right)
                     (\rd\theta^2 + \sin^2\theta \, \rd\phi^2) \right]
\nonumber  \\ & &
       \quad  {}+ 2 \frac{r^3}{9}
           \left(1-\frac{2\mu}{r}\right)\left(1-\frac{4\mu}{3r}\right)
       \calB^{(3)}_A \dt \dx^A.
\label{h3}
\end{eqnarray}
The part of ${}_3h_{ab}$ proportional to $\dot\calE_{ijk}$ or
$\dot\calB_{ijk}$ is of $\Or( r^4/\calR^4)$ and not required here.

At this level of approximation, the interactions of tidal forces with a small
black hole have no significant effect upon the motion of the hole. From the
analysis of Thorne and Hartle \cite{ThorneHartle85} the dominant tidal effect
upon the motion of a nonrotating object results from the coupling between the
external octupole moment of the geometry $\calE_{ijk}$ and the internal
quadrupole moment of the object $\calI_{jk}$; the resulting force is
\beq
    \mu a^i \sim \calE^i{}_{jk} \calI^{jk} ,
\eeq
equation (1.12) of reference \cite{ThorneHartle85}. For a Schwarzschild black
hole, $\calI_{jk}$ must result from the external quadrupole moment
$\calE_{jk}$. With dimensional analysis we conclude that this tidal
acceleration is no larger than
\beq
   a^i \sim \mu^4 \calE^i{}_{jk} \calE^{jk} \sim \mu^4/\calR^5 .
\eeq
This acceleration is much smaller than the $\Or(\mu/\calR^2)$ acceleration of
the self-force which is the focus of this manuscript.  Hence, we conclude
that for our purposes the tidal forces resulting from \Deqn{h2ev}-(\ref{h3})
exert no significant net force on the black hole.

\section{ A small black hole moving through a background geometry}
 \label{matching}

\subsection{Buffer region}
 \label{buffer}

In the previous section we treated the actual metric of a small black hole
moving through an external universe as the Schwarzschild metric being
perturbed by tidal forces with a small perturbation parameter $r/\calR$,
\begin{equation}
  g^\act_{ab} = g_{ab}^{\Text{\scriptsize Schw}} + {}_2h_{ab} + {}_3h_{ab}
   + \Or(r^4/\calR^4),
 \label{gabpertA}
\end{equation}
 The metric perturbations ${}_2h_{ab}$ and
${}_3h_{ab}$ are the dominant perturbations arising from the quadrupole and
octupole tidal forces and are given in \Deqns{h2ev}-(\ref{h3}).

In the buffer region $\mu \ll r \ll \calR$ the actual metric is described
equally well as the background metric being perturbed by the small mass $\mu$
with a perturbation parameter $\mu/r$. With THZ coordinates the background
metric is
\begin{eqnarray}
  g^0_{ab} &=&  \eta_{ab} + {}_2H_{ab} + {}_3H_{ab} + \Or( r^4/\calR^4)
\label{HsA}
\end{eqnarray}
and the actual metric is
\begin{eqnarray}
  g^\act_{ab} = g^0_{ab} + h^\mu_{ab} + h^{\mu^2}_{ab} + h^{\mu^3}_{ab} +
   \ldots
\end{eqnarray}
Each $h^{\mu^n}_{ab}$ is the part of the metric perturbation which is
proportional to $\mu^n$. These are obtained by a re-expansion of the results
of the previous section in terms of powers of the small parameter $\mu/r$.
Thus,
\begin{equation}
  h^\mu_{ab} \equiv {}_0h_{ab}^\mu + {}_2h_{ab}^\mu
          + {}_3h_{ab}^\mu + (\mu r^3/\calR^4) ,
\label{hS}
\end{equation}
where
\begin{equation}
  {}_0h^\mu_{ab}dx^a \dx^b = 2\frac{\mu}{r}(\dt^2 + \dr^2)
\label{h0}
\end{equation}
is the $\mu/r$ part of the Schwarzschild metric $g^\Text{\scriptsize
Schw}_{ab}$,
\bea
  {}_2h^\mu_{ab} \dx^a \dx^b &=&  4 \mu r\calE^{(2)} \dt^2
      - 2 \frac{2\mu r}{3} \calB^{(2)}_A \dt \dx^A
\nonumber\\ &&
  {} + 2 \frac{\mu r^2}{3} \dot\calE^{(2)} \dt \dr
         + 2 \frac{\mu r^2}{6} \dot\calB^{(2)}_A \dr \dx^A
\label{h2mu}
\eea
consists of the $\mu r/\calR^2$ and $\mu r^2/\calR^3$ parts of ${}_2h_{ab}$
from \Deqns{h2ev} and (\ref{h2od}), and ${}_3h_{ab}^\mu$ is the $\mu
r^2/\calR^3$ part of ${}_3h_{ab}$ in \Deqn{h3}
\begin{eqnarray}
 {}_3h^\mu_{ab} \dx^a \dx^b  &=&
      \frac{\mu r^2}{3}\calE^{(3)}
      \left[5 \dt^2 + \dr^2
             + 2r^2 (\rd\theta^2 + \sin^2\theta \rd\phi^2) \right]
\nonumber  \\ & &
       \quad  {} - 2 \frac{10\mu r^2}{27} \calB^{(3)}_A \dt \dx^A.
\label{h3muab}
\end{eqnarray}

\subsection{Asymptotic matching}
 \label{asymptotic}

To add a modest amount of formality to this analysis, we assume that the
background metric $g^0_{ab}$ with a geodesic $\Gamma$  has an expansion in
terms of THZ coordinates as in \Deqn{HsA}. We then consider a sequence of
metrics $g_{ab}(\mu)$ which are solutions of the vacuum Einstein equations
with a Schwarzschild black hole ``centered on $\Gamma$'' in the sense that
near the black hole the metric is approximately described as in
\Deqn{gabpertA}. The sequence is parameterized by $\mu \ll \calR$ with
$g_{ab}(0) = g^0_{ab}$. Our focus is on the behavior of $g_{ab}(\mu)$ in the
limit that $\mu \rightarrow 0$. This analysis falls under the purview of
singular perturbation theory \cite{BenderOrszag}: $g_{ab}(\mu)$ has an event
horizon if and only if $\mu \ne 0$; therefore, the exact metric for $\mu = 0$
differs fundamentally from a neighboring metric obtained in the limit $\mu
\rightarrow 0$.

In the buffer region $g_{ab}(\mu)$ is nicely illustrated in a fashion
introduced by Thorne and Hartle \cite{ThorneHartle85} as a sum of elements of
positive powers of the small parameters $\mu/r$ and $r/\calR$,
\begin{equation}
\begin{array}{cccccccccccccc}
\fl  g(\mu) & \sim
  & \eta       & \& & 0   & \& & {}_2H^\prime
               & \& & {}_3H^\prime
              & \& & {}_4H^\prime
 & \& & \cdots
 & =g^0
\\ \fl & \& & \mu/r      & \& & \mu /\calR      & \& & \mu r /\calR^2
               & \& & \mu r^2 /\calR^3
              & \& & \mu r^3 /\calR^4
 & \& & \cdots
 & = h^\mu
\\ \fl & \& & \mu^2/r^2  & \& & \mu^2 /r\calR   & \& & \mu^2  /\calR^2
               & \& & \mu^2 r /\calR^3
              & \& & \mu^2 r^2 /\calR^4
 & \& & \cdots
 & = h^{\mu^2}
\\ \fl & \& & \mu^3/r^3  & \& & \mu^3 /r^2\calR    & \& & \mu^3 /r\calR^2
               & \& & \mu^3 /\calR^3
              & \& & \mu^3 r /\calR^4
 & \& & \cdots
 & = h^{\mu^3}
\\ \fl & \& & \vdots && \vdots && \vdots && \vdots
              && \vdots
\\ \fl && \rule[.3em]{2em}{.05em} && \rule[.3em]{2em}{.05em}
   && \rule[.3em]{2em}{.05em} && \rule[.3em]{2em}{.05em}
              && \rule[.3em]{2em}{.05em}
\\ \fl  && g^{\Text{\scriptsize Schw}}
              &&  0   && {}_2h^\prime   && {}_3h^\prime
              && {}_4h^\prime
\end{array}
\label{tableau}
\end{equation}
where $\&$ means ``and an element of the form $\ldots$'' Starting with
$\el=0$, the $\el$th column in the tableau consists of elements which scale
as $(r/\calR)^\el$. Starting with $n=0$, the $n$th row consists of elements
which scale as $(\mu/r)^n$.
%% The tableau is not unique; part of any element can be moved diagonally, up
%% and to the left, to be absorbed into a dominating element with the same $r$
%% behavior.
In the $\mu/\calR\rightarrow 0$ limit, every non-zero element in the tableau
is larger than all elements below it in the same column, or to its right in
the same row.

The primes on the $H$'s in the top row work around a deficiency in our
notation: In section \ref{THZcoords} the prefix $2$ in ${}_2H_{ab}$ refers to
the multipole index $\el=2$. In the tableau, the prefix $2$ on
${}_2H^\prime_{ab}$ refers to the power of the order behavior,
$\Or(r^2/\calR^2)$. While ${}_2H_{ab}$ includes not only the quadrupole parts
proportional to $\calE_{ij}$ and  $\calB_{ij}$, which are $\Or(r^2/\calR^2)$,
but also the parts proportional to time derivatives of $\calE_{ij}$ and
$\calB_{ij}$, which are the order of a higher power of $r/\calR$. In the
tableau, the time derivative terms of ${}_\el H_{ab}$ are included in
${}_{\el+1}H^\prime_{ab}$ and columns further to the right.

Row $n$ is proportional to $\mu^n$ and is an expansion in the external
moments and in their time derivatives. Each element in the tableau is a
\textit{finite} combination of terms which scale with the same power of
$1/\calR$,
\beq
\fl  \mu^n r^{\el-n} /\calR^\el \sim  \left(\frac{\mu}{r}\right)^n r^\el
\left( \calE_\el  \quad\&\quad  {}^{(1)}\calE_{\el-1}
   \quad\&\quad  {}^{(2)}\calE_{\el-1}  \quad\&\quad  \cdots  \quad\&\quad
       {}^{(\el-2)}\calE_2 \right)
\eeq
The prefix superscript is the number of time derivatives, and
${}^{(p)}\calE_\el$ represents the even or odd parity $\el$ indexed STF
external multipole moment differentiated with respect to time $p$ times.
Thus, $\el$ is the largest external multipole index that contributes to any
element in column $\el$ or to ${}_\el h^\prime$.

At the outer edge of the buffer region, where $\mu/r \ll r/\calR$,
$g_{ab}(\mu)$ is approximately the background metric perturbed by $\mu$. In
this region, the top row of the tableau consists of the expansion of
$g^0_{ab}$ about $\Gamma$ in powers of $r/\calR$, contains no $\mu$
dependence and dominates the actual metric $g_{ab}(\mu)$.
 The sum of the top row is $g^0_{ab}$.

The $n=1$ row combines to give $h^\mu_{ab}$ which is the $\Or(\mu)$ metric
perturbation of $g^0_{ab}$. And the $n$th row combines to give the
$\Or(\mu^n)$ perturbation; higher order perturbation theory for the
background geometry is necessary to determine the $n>1$ rows.

At the inner edge of the buffer region, where $\mu/r \gg r/\calR$,
$g_{ab}(\mu)$ is approximately the Schwarzschild geometry perturbed by
background tidal forces. The $\el=0$ column of the tableau is simply an
expansion of the Schwarzschild geometry in powers of $\mu/r$, contains no
$\calR$ dependence and dominates the actual metric $g_{ab}(\mu)$.

The $\el=1$ column, linear in $r/\calR$, would be a dipole perturbation of
the Schwarzschild geometry. But there is no $r/\calR$ term in an expansion
about a geodesic. Consequently the top element of the $\el=1$ column is zero,
as are all elements of this column.

The top term in the $\el=2$ column, ${}_2H_{ab}^\prime$ represents the
external quadrupole tidal field. When this is combined with the rest of the
$\el=2$ column the result is ${}_2h_{ab}^\prime$, the entire quadrupole
perturbation of the black hole caused by tidal forces, in the time
independent approximation. ${}_2h^\prime_{ab}$ is given explicitly as in the
$\Or(1/\calR^2)$ terms of \Deqn{h2ev} and \Deqn{h2od}.

Similarly, the top term in the $\el=3$ column, ${}_3H_{ab}^\prime$ represents
the $\Or(r^3/\calR^3)$ external tidal field which distorts the black hole
creating ${}_3h_{ab}^\prime$, which is given as the $\Or(1/\calR^3)$ terms in
\Deqns{h2ev}-\Deqns{h3}. Thus, the top element of each column provides a
boundary condition for the equations which determine the resulting metric
perturbation of the black hole. Each column also satisfies appropriate
boundary conditions at the event horizon.

The analyses for ${}_\el h_{ab}^\prime$ up to $\el=3$ are straightforward
problems in linear perturbation theory of a Schwarzschild black hole. The
nonlinearity of the Einstein equations first appears in the elements of the
$\el=4$ column, which have some contributions from terms quadratic in the
$\el=2$ elements. Higher order perturbation theory for a black hole is
necessary to determine the $\el\ge4$ columns.

The actual metric is accurately approximated by $g_{ab}^{\Text{\scriptsize
Schw}}+{}_2h^\prime_{ab}+{}_3h^\prime_{ab}$ for $r \ll \calR$, and $g^0_{ab}$
is an accurate approximation of $g_{ab}(\mu)$ for $\mu \ll r$. In the buffer
region $\mu \ll r \ll \calR$ these approximations are
\begin{equation}
  g_{ab}^{\Text{\scriptsize Schw}} + {}_2h^\prime_{ab} + {}_3 h^\prime_{ab}
    = \eta_{ab} + {}_2H^\prime_{ab} + {}_3H^\prime_{ab} + \Or(\mu/r)
\label{gpert}
\end{equation}
and
\begin{equation}
  g^0_{ab} = \eta_{ab} + {}_2H^\prime_{ab} + {}_3H^\prime_{ab}
            + \Or(r^4/\calR^4) .
\label{g0}
\end{equation}
A demonstration of asymptotic matching \cite{BenderOrszag} requires a
\textit{matching zone}, within the buffer region, where the smallest
displayed term on the right hand side, ${}_3H^\prime_{ab} =
\Or(r^3/\calR^3)$, is simultaneously much larger than both remainder terms,
$\Or(\mu/r)$ and $\Or(r^4/\calR^4)$.
 The actual metric is accurately approximated by equation (\ref{gpert}) to the
``left'' of the matching zone, by equation (\ref{g0}) to the ``right'' of the
matching zone, and by $\eta_{ab} + {}_2H^\prime_{ab} + {}_3H^\prime_{ab}$
\textit{only} within the matching zone.

The matching zone is thus bounded by $\mu/r \ll r^3/\calR^3$ on the left and
by $r^4/\calR^4 \ll r^3/\calR^3$ on the right. These may be combined into
\beq
  (\mu\calR^3)^{1/4} \ll r \ll \calR,
\eeq
and this fits  within the buffer region because
\beq
  \mu \ll (\mu\calR^3)^{1/4} \ll r  \ll \calR,
                  \qquad \mu/\calR \rightarrow 0.
\eeq
This is the signature of a matched asymptotic expansion.

\section{Singular field $h^\sS_{ab}$}
 \label{Spart}

The Einstein tensor is the sum of terms consisting of the product of various
components of the metric and its inverse along with two derivatives. In the
buffer region, where $\mu \ll r \ll \calR$, an expansion of the Einstein
tensor $G_{ab}[g(\mu)]$ in positive powers of $\mu/r$ and $r/\calR$ may be
represented in a tableau similar to that for $g_{ab}(\mu)$ introduced in
section \ref{matching}.

In the expansion of $G_{ab}[g(\mu)]$ the terms of every power of $1/\calR$
which contain no dependence upon $\mu$ are each zero because $g^0_{ab}$ is
assumed to be a vacuum solution of the Einstein equations, $G_{ab}(g^0)=0$.
Similarly, all of the terms linear in $\mu$ must combine to yield
$E_{ab}(h^\mu) = -8\pi T_{ab}$, because $h^\mu_{ab}$ is a perturbative
solution of the Einstein equations with $T_{ab}$ representing a point mass.
The individual terms in $g_{ab}(\mu)$ which are linear in $\mu$ also form an
asymptotic expansion for $h^\mu$; these are the ${}_\el h^{\mu \prime}_{ab}$
terms in the $n=1$ row of the tableau for $g_{ab}(\mu)$.

In sections \ref{introduction} and \ref{perturbation} we discussed the actual
metric perturbation $h^\act_{ab}$ from a point mass moving through an
external geometry. The Hadamard form of the Green's function for the operator
$E_{ab}(h)$ provides a decomposition $h^\act_{ab} = h^\sS_{ab} + h^\R_{ab}$
in a neighborhood of $\Gamma$, where $E_{ab}(h^\sS)=-8\pi T_{ab}$. The
analysis of the Green's function yields an asymptotic expansion for
$h^\sS_{ab}$. The remainder $h^\R_{ab}$ is necessarily a vacuum solution of
$E_{ab}(h^\R)=0$ in a neighborhood of $\Gamma$ where an expansion for
$h^R_{ab}$ is regular. Thus, in the tableau for $g_{ab}(\mu)$, $h^R_{ab}$ is
$\Or(\mu)$. However, its regular behavior in a neighborhood of $\Gamma$
implies that it has no spatial dependence on a scale of $\Or(\mu)$, and that
it should properly be moved up in the tableau to be absorbed in the
definition of $g^0_{ab}$. This $\Or(\mu)$ change in $g^0_{ab}$ would affect
the $h^{\mu^n}_{ab}$ only for $n\ge2$. Further, the actual constructions of
${}_2h^\prime_{ab}$ and ${}_3h^\prime_{ab}$, resulting in equations
(\ref{h2ev})-(\ref{h3}), do not appear to allow for the inclusion of any such
regular part, except in the top row.

The possibility that $h^\R_{ab}$ when promoted to the top row, might contain
a dipole part in the $\el=1$ column is discussed in section \ref{Rpart}.

With no clear proof at hand, we thus provide the conjecture that the ${}_\el
h^{\mu\prime}_{ab}$ are the terms in an asymptotic expansion for $h^\sS_{ab}$
and, therefore, that
\beq
     h^\sS_{ab} = h^\mu_{ab}
\eeq
and that $h^\R_{ab}$ is included in the top row of the tableau
(\ref{tableau}).  We have verified that ${}_0h^{\mu}_{ab}$ and
${}_2h^{\mu}_{ab}$ (in the Lorenz gauge) are equivalent to the first two
terms in the expansion of $h^\sS_{ab}$ via the Hadamard form of the Green's
function. Further, the
 ${}_\el h^{\mu\prime}_{ab}$ have no dipole $\el=1$ component at $\Or(\mu)$ which
could effect the world line $\Gamma$ at a level of interest in a self-force
calculation.

In the next two sections the effect of coordinate choices on the form of
$h^\sS_{ab}$ are discussed. First, a change in the locally-inertial
coordinates appears as a gauge transformation of the Schwarzschild metric
being perturbed by the external tidal fields. Second, an $\Or(\mu)$
coordinate change appears as a gauge transformation of the background metric
being perturbed by a point mass $\mu$.

\subsection{Coordinate transformations of the locally inertial coordinates}
 \label{localgauge}

The convenient THZ coordinate system is used in sections \ref{slow} and
\ref{matching} to determine the leading terms ${}_0h^{\mu}_{ab}$,
${}_2h^{\mu\prime}_{ab}$ and ${}_3h^{\mu\prime}_{ab}$ in an expansion of
$h_{ab}^\sS$. But, if $h_{ab}^\sS$ is to play a fundamental role in radiation
reaction and self-force analyses then the definition of $h_{ab}^\sS$ should
certainly not be wed to any particular locally-inertial coordinate system.

In this section we examine the change in the description of $h^\sS_{ab}$
under a change of locally-inertial coordinates. The next section describes
how an $\Or(\mu r^\el/\calR^\el)$ gauge transformation of the perturbed
Schwarzschild metric changes the form of $h^\sS_{ab}$ while remaining with
the same locally-inertial coordinates.

For the ``inner'' perturbation problem of the matched asymptotic expansions,
the external tidal fields are considered a perturbation of the Schwarzschild
geometry. From this perspective a change from one locally-inertial coordinate
system to another appears as a gauge transformation of the perturbed
Schwarzschild metric.

A second  locally-inertial coordinate system is defined by
\beq
  y^a = x^a + \lambda^a{}_{ijk}x^i x^j x^k + \Or(r^4/\calR^3),
\eeq
where $\lambda^a{}_{ijk}$ is an $\Or(1/\calR^2)$  constant, such as in
equation (\ref{thzfn}) which relates Fermi normal to THZ coordinates.
 For the perturbed Schwarzschild metric this appears as a gauge transformation
with a gauge vector $\xi^a=\lambda^a{}_{ijk}x^i x^j x^k + \Or(r^4/\calR^3)$.
Under such a change in coordinates the description of $h_{ab}^\sS$ changes in
two different ways: the functional dependence upon coordinate position
changes and the components of the tensor change. Let the components in the
$y$ coordinate system be denoted by a prime.
 For a fixed coordinate position $\kappa^c$,
\begin{eqnarray}
  h^\sS_{a^\prime b^\prime}\big|_{y^c=\kappa^c}
      & = & \big( h^\sS_{ab}\big|_{x^c
         = \kappa^c} - \xi^c \partial_c h^\sS_{ab}\big)
            \frac{\partial x^a}{\partial y^{a^\prime}}
            \frac{\partial x^b}{\partial y^{b^\prime}}
% \nonumber\\ &&
   {} + \Or(\mu r^2/\calR^3),
\end{eqnarray}
which, when expanded out, is
\begin{equation}
  h^\sS_{a^\prime b^\prime}
       = h^\sS_{ab} - \xi^c \partial_c h^\sS_{ab}
           - 2 h^\sS_{c(a} \partial_{b)} \xi^c
           + \Or(\mu r^2/\calR^3).
\end{equation}
The left hand side is evaluated at $y^c = \kappa^c$ and the right hand side
at $x^c = \kappa^c$. In terms of the Lie derivative $\Lie$, the descriptions
of the single tensor field $h^\sS_{ab}$ in two different locally-inertial
coordinate systems are related by
\begin{equation}
  h^{S}_{a^\prime b^\prime}  =  h ^{S}_{ab} - \Lie_\xi \, h^\sS_{ab}
                             + \Or(\mu r^2/\calR^3).
\end{equation}
Now, $h_{ab}^\sS = {}_0h_{ab}^\mu + {}_2h_{ab}^\mu + \Or(\mu r^2/\calR^3)$,
as in \Deqn{hS}, and ${}_0h^\mu_{ab} = \Or(\mu/r)$ in any locally-inertial
coordinates. Thus, the change in $h^\sS_{ab}$ is most naturally assigned to
${}_2h^\mu_{ab}$,
\begin{equation}
 {}_2h^{\mu\Text{\scriptsize new}}_{ab}
     =  {}_2h^{\mu\Text{\scriptsize old}}_{ab}
              - \Lie_\xi \, {}_0h^\mu_{ab} + \Or(\mu r^2/\calR^3) .
\label{delh2mu}
\end{equation}

This description of the change in the ${}_2h^\mu_{ab}$ part of $h^\sS_{ab}$
is consistent with the related gauge transformation of the $\ell = 2$ metric
perturbation, ${}_2h_{ab} = {}_2H_{ab} + {}_2h_{ab}^\mu +
\Or(\mu^2/\calR^2)$, of the Schwarzschild geometry
\begin{equation}
  {}_2h^{\Text{\scriptsize new}}_{ab}
  = {}_2h^{\Text{\scriptsize old}}_{ab} - \Lie_\xi
  g^{\Text{\scriptsize Schw}}_{ab}
          + \Or(r^3/\calR^3).
\label{delh2}
\end{equation}
The leading terms of this for large $r$ are
\begin{eqnarray}
\fl  {}_2H^{\Text{\scriptsize new}}_{ab} + {}_2h^{\mu\Text{\scriptsize
new}}_{ab} &=&
                 {}_2H^{\Text{\scriptsize old}}_{ab}
               + {}_2h^{\mu\Text{\scriptsize old}}_{ab}
    {} - \Lie_\xi   (\eta_{ab} + {}_0h^\mu_{ab})
    {}  + \Or(r^3/\calR^3,\, \mu r^2/\calR^3).
\label{delH2}
\end{eqnarray}
These are naturally apportioned as
\begin{eqnarray}
  {}_2H^{\Text{\scriptsize new}}_{ab}
  &=& {}_2H ^{\Text{\scriptsize old}}_{ab} - \Lie_\xi \eta_{ab}
          + \Or(r^3/\calR^3)
\nonumber \\ & = &
       {}_2H ^{\Text{\scriptsize old}}_{ab} - 2\nabla_{(a} \xi_{b)}
          + \Or(r^3/\calR^3)
%\label{delH}
\end{eqnarray}
and
\begin{equation}
  {}_2h^{\mu\Text{\scriptsize new}}_{ab}
  =  {}_2h ^{\mu\Text{\scriptsize old}}_{ab}
            - \Lie_\xi \,{}_0h^\mu_{ab}
          + \Or(\mu r^2/\calR^3).
\label{delh}
\end{equation}
A comparison of \Deqns{delh2mu} and (\ref{delh}) reveals the consistency of
the description of $h^\sS_{ab}$ as a single tensor field, which in any normal
coordinate system is approximated by ${}_0h_{ab}^\mu + {}_2h_{ab}^\mu +
{}_3h_{ab}^\mu + \Or(\mu r^3/\calR^4) $ for $\mu \ll r \ll \calR$.

An $\Or(r^4/\calR^3)$ transformation changes ${}_3h_{ab}^\mu$ in a similar
way.

\subsection{Transformation of  $h^\sS_{ab}$ to the Lorenz gauge}
  \label{lorenzgauge}

The convenient Regge-Wheeler gauge was used, with the THZ coordinates, in
sections \ref{slow} and \ref{matching} to determine the leading terms
${}_0h^{\mu}_{ab}$, ${}_2h^{\mu\prime}_{ab}$ and ${}_3h^{\mu\prime}_{ab}$ in
an expansion of $h_{ab}^\sS$. But, if $h_{ab}^\sS$ is to play a fundamental
role in radiation reaction and self-force analyses then the definition of
$h_{ab}^\sS$ should certainly not be wed to any particular gauge choice.

This section gives an example of an $\Or(\mu r^\el/\calR^\el)$ gauge
transformation, for $\el=0$ and $2$, of the perturbed background metric
$g^0_{ab}$ which changes the form of $h^\sS_{ab}$ while remaining with the
same locally-inertial coordinates. The previous subsection describes how the
description of $h^\sS_{ab}$ changes under a change of locally-inertial
coordinates.

$h^\sS_{ab}$ is given above in \Deqn{hS}-(\ref{h2mu}) in the Regge-Wheeler
gauge. To transform the ${}_2h^{\mu\prime}_{ab}$ part of this into the Lorenz
gauge, the gauge vector is
%    \beq
%      \xi^i = - \mu \frac{x^i}{r} (1  + \calE_{jk}x^j x^k)
%                 + 2\mu r \calE^i{}_j x^j .
%    \eeq
%  \textBF{Replace the above with (checked this, especially the units):}
\beq
  \xi^a = - \mu (1  - r^2 \calE^{(2)}) \delta^a_r
             + \mu r^2 \calE^{(2)}_B \sigma^{Ba} .
\eeq
The Lorenz gauge has
\beq
  {}_2h^{\mu\prime}_{ab}(\lz) = {}_2h^{\mu\prime}_{ab}(\rw)
                              - \nabla_a \xi_b - \nabla_b \xi_a ,
\eeq
where the metric being perturbed is $g^0_{ab}$, and $\nabla_a$ is its
covariant derivative operator. This results in
\bea
 \fl {}_2h^{\mu\prime}_{ab}(\lz) \dx^a \dx^b &=&
    \frac{2\mu}{r} \left[ (1+r^2\calE^{(2)}) \dt^2
      {} + (1 - 3 r^2 \calE^{(2)} ) \dr^2
         + (1 - r^2 \calE^{(2)} ) \sigma_{AB}\dx^A \dx^B \right]
\nonumber\\ \fl &&
       {} - 4 \mu r \calE^{(2)}_A \dr \dx^A
        - 2 \mu r \calE^{(2)}_{AB} \dx^A \dx^B
\nonumber\\ \fl &&
       {} + 2 \frac{\mu}{3r}  \calB^{(2)}_A  \dt \dx^A
          + \Or(\mu r^2/\calR^3)  .
\label{hlz}
\eea

For completeness, the trace of ${}_2h^{\mu\prime}_{ab}(\lz)$ is
\beq
 (\eta^{ab} - {}_2H^{ab}) {}_2h^{\mu\prime}_{ab}(\lz)
         = 4\mu/r +\Or(\mu r^2/\calR^3),
\eeq
and the trace-reversed
 ${}_2\bar{h}^{\mu\prime}_{ab} \equiv  {}_2h^{\mu\prime}_{ab}
  - \frac{1}{2} g^0_{ab} g_0^{cd}{}_2h^{\mu\prime}_{cd} $
is
\bea
 {}_2\bar{h}^{\mu\prime}_{ab}(\lz) \dx^a \dx^b &=&
    \frac{4\mu}{r} (1+ r^2\calE^{(2)} ) \dt^2
     {} - 4 \mu r \calE^{(2)} \dr^2
\nonumber\\ &&
     {} - 4 \mu r \calE^{(2)}_A \dr \dx^A
        - 2 \mu r \calE^{(2)}_{AB} \dx^A \dx^B
\nonumber\\ &&
       {} - 2 \frac{r\mu}{3} \calB^{(2)}_A \dt \dx^A ,
\eea
which satisfies the Lorenz gauge condition.
\beq
  \nabla^a {}_2\bar{h}^{\mu\prime}_{ab}(\lz) = \Or(\mu r/\calR^3) .
\label{lzgauge}
\eeq

Equation (\ref{hlz}) gives ${}_2h^{\mu\prime}_{ab}$ in the Lorenz gauge with
THZ coordinates. From the perspective of the background metric $g^0_{ab}$, a
change from THZ to Fermi normal coordinates, as described in the previous
section, would preserve the covariant condition (\ref{lzgauge}) for the
Lorenz gauge and provide $h^\sS_{ab}(\lz)$ in Fermi normal coordinates.

\section{Regular field $h^\R_{ab}$}
\label{Rpart}

In a self-force application, it is first required to find the actual metric
perturbation $h^\act_{ab}$ for a point mass $\mu$ moving along a geodesic
$\Gamma$ of the background spacetime $g^0_{ab}$. In many cases $h^\act_{ab}$
will be the retarded metric perturbation. However, we prefer to leave the
choice of boundary conditions general.

From the expansion of $g^0_{ab}$ about $\Gamma$, as in \ref{inertial}, the
first few terms of an asymptotic expansion for $h^\sS_{ab}$ is determined as
in \ref{matching}. The \textit{regular remainder} is defined by
\beq
  h^\R_{ab} \equiv h^\act_{ab} - h^\sS_{ab}
\label{hR}
\eeq
in a neighborhood of $\Gamma$ where $E_{ab}(h^\R) = 0$. $h^R_{ab}$ does not
change over an $\Or(\mu)$ length scale, so it is natural to combine
$h^R_{ab}$ with $g^0_{ab}$ in the top row of the tableau of \ref{matching}.
Then the condition that the dipole term of the top row is zero is equivalent
to the condition that $\Gamma$ is actually a geodesic of $g^0_{ab} +
h^\R_{ab}$.

From a different perspective, if $h^R_{ab}$ is left in the $\mu^1$ row, and
if $\Gamma$ is not a geodesic of $g^0_{ab} + h^\R_{ab}$, then $h^\R_{ab}$
necessarily has a dipole part in its expansion about $\Gamma$. This implies
that the gravitational field of $\mu$ is not centered upon $\Gamma$. The act
of adjusting $\Gamma$ to remove the dipole field and to accurately track the
center of the gravitational field of $\mu$ is equivalent to requiring that
$\Gamma$ be a geodesic of $g^0_{ab}+h^\R_{ab}$. This act of adjustment is
also equivalent to performing a gauge transformation to the perturbed
Schwarzschild geometry that removes the dipole field.

Thus the consistency of the matched asymptotic expansions implies that,
indeed, the $\Or(\mu)$ correction to geodesic motion for an infinitesimal
black hole has the motion being geodesic in $g^0_{ab}+h^\R_{ab}$.

In an actual calculation, an exact expression for $h^\sS_{ab}$ is usually not
available. It is only necessary that $g^0_{ab}+h^\R_{ab}$ be $C^1$ in order
to determine a geodesic, and this $C^1$ requirement can be met as long as the
approximation for $h^\sS_{ab}$ includes at least ${}_0h^{\mu\prime} +
{}_2h^{\mu\prime}$.  The next term is ${}_3h^{\mu\prime} = \Or(\mu
r^2/\calR^3)$, and its derivative necessarily vanishes on $\Gamma$ where
$r=0$.  Thus, calculations of the self-force will be successful as long as
the monopole and quadrupole terms of the asymptotic expansion for
$h^\sS_{ab}$ are included in the evaluation of $h^\R_{ab}$ via \Deqn{hR}.
Nevertheless, including the higher order terms of $h^\sS_{ab}$, results in
the approximation for $h^\R_{ab}$ being more differentiable. In a
calculation, usually $h^\act_{ab}$ is determined as a sum over modes with
$h^\R_{ab}$ being decomposed in terms of the same modes. In determining the
self-force, the more differentiable the $h^\R_{ab}$ is, the more rapidly the
sum over modes converges. The use of higher order terms in an approximation
for $h^\sS_{ab}$ can have dramatic effects on the ultimate accuracy of a
self-force calculation \cite{DetMessWhiting03}.

If we have the actual $\Or(\mu)$ metric perturbation $h^\act_{ab}$ for a
point mass, then the asymptotic matching provides an approximation for the
geometry of a small black hole moving in the external geometry $g_{ab}(\mu)$
in the limit that $\mu/\calR \rightarrow 0$. The approximation extends
throughout the entire external spacetime down to the event horizon. Further,
the approximation is revealed to be uniformly valid by the concise
description of the matched geometry as
\begin{eqnarray}
 \fl g_{ab}(\mu) &=&(g^0_{ab}+h_{ab}^\act)
    + (g_{ab}^{\Text{\scriptsize Schw}}
    + {}_2h_{ab}^\prime + {}_3h_{ab}^\prime)
\nonumber \\
  \fl  &&{}- (\eta_{ab} + {}_2H_{ab}^\prime + {}_3H_{ab}^\prime
        + {}_0h^{\mu\prime}_{ab}
        + {}_2h^{\mu\prime}_{ab} + {}_3h^{\mu\prime}_{ab})
    + \Or(\mu^2/\calR^2),
                    \; \mu/\calR\rightarrow 0.
\label{gmatch}
\end{eqnarray}
The combination  $g^0_{ab}+h_{ab}^\act$ includes all terms in the top two
rows of the tableau but extends outside the buffer region to include the
entire external spacetime. The combination $g_{ab}^{\Text{\scriptsize Schw}}
+ {}_2h_{ab}^\prime + {}_3h_{ab}^\prime$ is the left four columns in the
tableau. The remaining terms keep the entire expression from double-counting
the elements in the upper left corner. The dominant term from the tableau
which is not included here is $\Or(\mu^2 r^2/\calR^4)$ which gives the
$\Or(\mu^2/\calR^2)$ remainder for this uniformly valid approximation for the
matched metric in the limit $\mu/\calR\rightarrow 0$.

\section{Gauge issues}
\label{gauge}

\subsection{Gauge transformations}

In perturbation analyses of general relativity \cite{Sachs64,
StewartWalker74, Bardeen80}, one considers the difference in the actual
metric ${g}^\act_{ab}$ of an interesting, perturbed spacetime and the
abstract metric ${{g}}^0_{ab}$ of some given, background spacetime. The
difference
\beq
  h_{ab} = {g}^\act_{ab} - g^0_{ab}
 \label{hdef}
\eeq
is assumed to be infinitesimal, say $\Or(h)$. Typically, one determines a set
of linear equations which govern $h_{ab}$ by expanding the Einstein equations
through $\Or(h)$. The results are often used to resolve interesting issues
concerning the stability of the background, or the propagation and emission
of gravitational waves by a perturbing source.

However, \Deqn{hdef} is ambiguous: The metrics ${g}^\act_{ab}$ and
${{g}}^0_{ab}$ are given on different manifolds. For a given event on one
manifold at which corresponding event on the other manifold is the
subtraction to be performed? Usually a coordinate system common to both
spacetimes induces an implicit mapping between the manifolds and defines the
subtraction. Yet, the presence of the perturbation allows ambiguity. An
infinitesimal coordinate transformation of the perturbed spacetime
\beq
  \prim{x}^a = x^a + \xi^a , \quad \Text{where} \quad \xi^a = \Or(h),
 \label{gaugetransformation}
\eeq
not only changes the components of a tensor at $\Or(h)$, in the usual way,
but also changes the mapping between the two manifolds in \Deqn{hdef}. After
the transformation \Deqn{gaugetransformation},
\bea % \prim
  h^\Text{\scriptsize new}_{ab}
    &=&
     \left({{g}}^0_{cd} + h^\Text{\scriptsize old}_{cd}\right)
         \frac{\partial x^c}{\partial \prim x^a}
         \frac{\partial x^d}{\partial \prim x^b}
   - \left( {{g}}^{0}_{ab}
       + \xi^c \frac{\partial {{g}}^{0}_{ab}}{\partial x^c}\right) .
\eea
The $\xi^c$ in the last term accounts for the $\Or(h)$ change in the event of
the background used in the subtraction. After an expansion, this provides a
new description of $h_{ab}$
\bea
   h^\Text{\scriptsize new}_{ab} &=& h^\Text{\scriptsize old}_{ab}
     - {{g}}^0_{cb} \frac{\partial \xi^c}{\partial x^a}
     - {{g}}^0_{cb} \frac{\partial \xi^d}{\partial x^b}
     - \xi^c \frac{\partial {{g}}^0_{ab}}{\partial x^c}
\nonumber \\ &=&  h^\Text{\scriptsize old}_{ab}
     -\Lie_{\xi} {{g}}^0_{ab}
  =  h^\Text{\scriptsize old}_{ab} - 2 \nabla_{(a} \xi_{b)}
\label{hprime}
\eea
through $\Or(h)$; the symbol $\Lie$ represents the Lie derivative and
$\nabla_{a}$ is the covariant derivative compatible with $g^0_{ab}$.

The action of such an infinitesimal coordinate transformation is called a
\textit{gauge transformation} and does not change the actual perturbed
manifold, but it does change the coordinate description of the perturbed
manifold.

A similar circumstance holds with general coordinate transformations. A
change in coordinate system creates a change in description. But, general
covariance dictates that actual physical measurements must be describable in
a manner which is invariant under a change in coordinates. Thus, one usually
describes physically interesting quantities strictly in terms of geometrical
scalars which, by nature, are coordinate independent.

In a perturbation analysis any physically interesting result ought to be
describable in a manner which is gauge invariant.

\subsection{Gauge invariant quantities}
\label{gaugeinvariantquantities}

Gauge invariant quantities appear to fall into a few different categories.

The change in any geometrical quantity under a gauge transformation is
determined by the Lie derivative of that same quantity on the background
manifold. This is demonstrated for the gauge transformation of a metric
perturbation in \Deqn{hprime}. We also used this fact to describe the change
in $h^\sS_{ab}$ under gauge transformations in sections \ref{localgauge} and
\ref{lorenzgauge}. Thus, if a geometrical quantity vanishes in the
background, but not in the perturbed metric, then it will be gauge invariant.
Examples include the Newman-Penrose scalars $\Psi_0$ and $\Psi_4$ which
vanish for the Kerr metric. In the perturbed Kerr metric $\Psi_0$ and
$\Psi_4$ are non zero, gauge invariant and the basis for perturbation
analyses of rotating black holes. A second example has the background metric
being a vacuum solution of the Einstein equations, so its Ricci tensor
$R_{ab}$ vanishes. The Ricci tensor of a perturbation of this metric is then
unchanged by a gauge transformation. This is directly demonstrated in
\ref{gaugeinvariance}.

Some quantities which are associated with a symmetry of the perturbed
geometry are gauge invariant. For example a geodesic of a perturbed
Schwarzschild metric, where the perturbation is axisymmetric with Killing
field $k^a$, has a constant of motion $k^a u^b (g^0_{ab}+h_{ab})$ which is
gauge invariant.

Another symmetry example involves the Schwarzschild geometry with an
arbitrary perturbation. It is a fact that a gauge transformation can always
be found, such that the resulting $h_{ab}$ has the components
$h_{\theta\theta}$, $h_{\theta\phi}$ and $h_{\phi\phi}$ all equal to zero. In
this gauge, the surfaces of constant $r$ and $t$ are geometrical two-spheres,
even while the manifold as whole has no symmetry. The area of each two-sphere
can be used to define a radial scalar field $R$ which is constant on each of
these two-spheres. This scalar field on the perturbed Schwarzschild manifold
is independent of gauge. However, its coordinate description in terms of the
usual $t$, $r$, $\theta$, $\phi$ coordinates does change under a gauge
transformation.  We find a use for this gauge invariant scalar field in
section \ref{SFcircular}.

Quantities which are carefully described by a physical measurement are gauge
invariant. For example, the acceleration of a world line could be measured
with masses and springs by an observer moving along a world line in a
perturbed geometry. The magnitude of the acceleration is a scalar and is
gauge invariant. If the world line has zero acceleration, then it is a
geodesic. Therefore, a geodesic of a perturbed metric remains a geodesic
under a gauge transformation even while its coordinate description changes by
$\Or(h)$.

The mass and angular momentum are other gauge invariant quantities which
might be measured by distant observers in an asymptotically flat spacetime. A
small mass orbiting a larger black hole perturbs the black hole metric and
emits gravitational waves. The gravitational waveform measured at a large
distance is also gauge invariant.

\subsection{Gauge transformations and the gravitational self-force}
 \label{GaugeTransGSF}
We understand that a point mass moves along a world line of a background
metric $g^0_{ab}$ and causes a metric perturbation $h^\act_{ab}$, which may
be decomposed into $h^\sS_{ab}$ and $h^\R_{ab}$. The gravitational self-force
makes the world line be a geodesic of $g^0_{ab}+h^\R_{ab}$. This world line
is equivalently described in terms of the background metric and its
perturbation by
\begin{equation}
  u^b \nabla_b u^a = - (g_0^{ab}+ u^a u^b) u^c u^d
     (\nabla_{c} h^\R_{db} - \frac12 \nabla_b h^\R_{cd}) = \Or(\mu/\calR^2)
 \label{gravforce}
\end{equation}
where the covariant derivative and normalization of $u^a$ are compatible with
$g^0_{ab}$.

Given this world line, let $\xi^a$ be a differentiable vector field which is
equal, on the world line, to the $\Or(\mu)$ displacement back to the geodesic
of $g^0_{ab}$ along which the particle would move in the absence of
$h^\R_{ab}$; otherwise $\xi^a$ is arbitrary. Such a $\xi^a$ generates a gauge
transformation for which the right hand side of \Deqn{gravforce} is zero when
evaluated with the new $h_{ab}$ \cite{BarackOri01}. With the new $h_{ab}$
there is no ``gravitational self-force'', and the coordinate description of
the world line is identical to the coordinate description of a geodesic of
$g^0_{ab}$. With or without the gauge transformation, an observer moving
along this world line would measure no acceleration and would conclude that
the world line is a geodesic of the perturbed metric.

This example shows that simple knowledge of the gravitational self-force, as
defined in terms of the right hand side of \Deqn{gravforce}, is not a
complete description of any physically interesting quantity.

With this same example, after a time $T$ the gauge vector $\xi \sim T^2 \dot
u \sim T^2 \mu/\calR^2$, and as long as $T \lesssim \calR$ the gauge vector
$\xi^a$ remains small. However, when $T\sim \calR \sqrt{\calR/\mu}$, which is
much larger than the dynamical timescale $\calR$, the gauge vector $\xi \sim
\calR$ and can no longer be considered small. Thus, a gauge choice which
cancels the coordinate description of the self-force necessarily fails after
a sufficiently long time. Mino \cite{Mino03} takes advantage of this fact in
his proposal to find the cumulative, gravitational self-force effect on the
Carter constant for eccentric orbits around a rotating black hole.

\section{An example: self force on circular
              orbits of the Schwarzschild metric}
\label{example}

The introduction described the Newtonian problem of a small mass $\mu$ in a
circular orbit of radius $R$ about a much larger mass $M$. The analysis
results in the usual $\Or(\mu/M)$ reduced mass effect on the orbital
frequency $\Omega$ given in \Deqn{newtonianOm2}. Reference
\cite{DetPoisson04} has a thorough introduction to the mechanics of this
self-force calculation and gives a detailed discussion of this elementary
problem using the same language and style which is common for the
relativistic gravitational self-force. This includes elementary expressions
for the  S and R-fields of the Newtonian gravitational potential with
descriptions of their decompositions in terms of spherical harmonics.

The extension of this Newtonian problem to general relativity is perhaps the
simplest, interesting example of the relativistic gravitational self-force.
Thus, we focus on a small mass $\mu$ in a circular geodesic about a
Schwarzschild black hole of mass $M$, and we describe each of the steps
necessary to obtain physically interesting results related to the
gravitational self-force.

\subsection{Mode sum analysis}

Metric perturbations of Schwarzschild have been thoroughly studied since
Regge and Wheeler \cite{ReggeWheeler, Zerilli}. Both $T_{ab}$ and $h_{ab}$
are fourier analyzed in time, with frequency $\omega$, and decomposed in
terms of tensor spherical harmonics, with multipole indices $\el$ and $m$.
Linear combinations of the components of $h^{\lm,\omega}_{ab}$ satisfy
elementary ordinary differential equations which are easily numerically
integrated. With the periodicity of a circular orbit, only a discrete set
frequencies $\omega_m = -m\Omega$ appear.

We assume, then, that $h^{\lm,\omega}_{ab}(r)$ can be determined for any
$\el$ and $m$. The sum of these over all $\el$ and $m$ then constitutes
$h^\act_{ab}$, and this sum will be divergent if evaluated at the location of
$\mu$.

The next task is to determine $h^\sS_{ab}$. The THZ coordinates, including
$\Or(r^4/\calR^3)$ terms, for a circular orbit of Schwarzschild are given in
reference \cite{DetMessWhiting03}. Equations (\ref{h0})-(\ref{h3muab}) give
an approximation for $h^\sS_{ab}$ in THZ coordinates with a remainder of
$\Or(\mu r^3/\calR^4)$.

We follow the mode-sum regularization procedure pioneered by Barack and Ori
\cite{BarackOri00, Barack00, BarackOri02, BarackOri03} and Mino, Sasaki and
Tanaka \cite{Mino02, BMNOS02} and followed up by others
\cite{DetMessWhiting03, Messaritaki03, Lousto00, DongHoon04}. In this
procedure, the multipole moments of the S-field are calculated and referred
to as \textit{regularization parameters}. The sum of these moments diverges
when evaluated at the location of $\mu$, but each individual moment is
finite. Importantly, the S-field has been constructed to have precisely the
same singularity structure at the particle as the actual field has. Thus the
difference in these moments gives a multipole decomposition of the regular
R-field.  Schematically, this procedure gives
\beq
  h^\R_{ab} = \sum_{\lm,\omega} h^{\R(\lm,\omega)}_{ab}
   = \sum_{\lm,\omega} \left[ h^{\act(\lm,\omega)}_{ab}
        - h^{\sS(\lm,\omega)}_{ab} \right]
\label{modesum}
\eeq
for the regular field.

We note that the sum over modes of a decomposition of a $C^\infty$ function
converges faster than any power of $\el$. And, the less the differentiability
of the function then the slower the convergence of its mode sum. Exact values
for $h^{\sS(\lm,\omega)}_{ab}$ would then give rapid convergence of the sum
yielding a $C^\infty$ representation of $h^\R_{ab}$. However, the
approximation for $h^\sS_{ab}$ in (\ref{h0})-(\ref{h3muab}) has an $\Or(\mu
r^3/\calR^4)$ remainder which is necessarily only $C^2$. This immediately
puts a limitation on the rate of convergence of any mode sum for $h^\R_{ab}$.
Further, $h^\sS_{ab}$ is only defined in a neighborhood of $\mu$. Whereas a
decomposition in terms of spherical harmonics requires a field defined over
an entire two-sphere. It is important that the extension of $h^\sS_{ab}$ over
the two-sphere is $C^\infty$ everywhere, except at $\mu$, to insure rapid
convergence of the mode sum. This ambiguity for $h^\sS_{ab}$, away from
$\mu$, highlights an important fact: the value of any individual multiple
moment $h^{\sS(\lm,\omega)}_{ab}$ or $h^{\R(\lm,\omega)}_{ab}$ is inherently
ill defined. Only a sum over modes, such as in \Deqn{modesum}, might have
physical meaning.

\subsection{Gauge issues}

A vexing difficulty with equation \Deqn{modesum} revolves around gauge
transformations. What assurance do we have that the singularity structure of
$h^\sS_{ab}$ truly matches the singularity structure of $h^\act_{ab}$? For
example $h^\sS_{ab}$ is often described in the Lorenz gauge which is well
behaved by most standards, whereas $h^\act_{ab}$ is most easily calculated in
the Regge-Wheeler gauge which often entails discontinuities in components of
$h_{ab}$.

A gauge transformation does not change the relationship
\beq
      E_{ab}(h^\act) =  E_{ab}(h^\sS) + E_{ab}(h^\R) = - 8\pi T_{ab} .
\eeq
But it also does not dictate how to apportion a gauge transformation for
$h^\act_{ab}$ between $h^\sS_{ab}$ and $h^\R_{ab}$. In a neighborhood of the
particle $h^\R_{ab}$ is known to be a solution of $E_{ab}(h^\R)=0$, but a
gauge transformation generates a homogeneous solution $-2\nabla_{(a}\xi_{b)}$
to the same equation, thus $h^\R_{ab}$ can determined only up to a gauge
transformation. Even a distribution-valued gauge transformation might be
allowed allowed because \ref{singulargauge} shows that $E_{ab}(\nabla\xi) =
0$, in a distributional sense, even in that extreme case. Thus it is expected
that $h^\R_{ab}$ calculated from \Deqn{modesum} might have a
non-differentiable part resulting from a singular gauge difference between
$h^\act_{ab}$ and $h^\sS_{ab}$.

My personal perspective on this situation is reassuring, at least to me, but
certainly not rigorous. I have considered about a half-dozen different
gravitational self-force problems involving a small point mass orbiting a
much larger black hole. In each problem the goal is the calculation of an
interesting, well-defined gauge invariant quantity. For each of these, the
natural formulation of the problem shows that there are ways to define and to
calculate the relevant quantities which are not deterred by a difference in
gauge between $h^\sS_{ab}$ and $h^\act_{ab}$, even if the difference involves
a distribution-valued gauge vector. It appears as though a particularly
odious gauge choice might exist for a specific problem, which might interfere
with a calculation. However, none of a wide variety of natural choices for a
gauge have this difficulty for the problems that I have examined.
Specifically, the example in this section appears to avoid any difficult
gauge issue.

\subsection{Geodesics of the perturbed Schwarzschild metric}
\label{SFcircular}

A particle of mass $\mu$ in a circular orbit about a black hole perturbs the
Schwarzschild metric by $h^\act_{ab} = \Or(\mu)$. The circumstances dictate
boundary conditions with no gravitational radiation incoming from infinity or
outgoing from the event horizon.

The dynamical timescale for a close orbit is $\Or(M)$ and much shorter than
the dynamical timescale $\Or(M^2/\mu)$ for radiation reaction to have a
significant effect upon the orbit. Thus the particle will orbit many times
before its orbital frequency changes appreciably. Under these conditions, the
perturbed metric appears unchanging in a coordinate system that rotates with
the particle. Thus, for times much less than the radiation reaction time,
there is a Killing vector $k^a$,
\beq
  \Lie_k (g^0_{ab}+h^\act_{ab}) = 0 ,
\label{killing}
\eeq
whose components in the usual Schwarzschild coordinates are
\beq
  k^a \frac{\partial}{\partial x^a} = \frac{\partial}{\partial t}
     + \Omega \frac{\partial}{\partial \phi} .
\label{orbitk}
\eeq

Let an observer on the particle be equipped with a flashlight which he holds
pointing in the plane of the orbit at a fixed orientation with respect to the
tangent to the orbit. In other words the orientation of the flashlight is Lie
derived by $k^a$ and the beam of light sweeps around the equatorial plane
once every orbit. A distant observer, in the equatorial plane, measures the
time $\Delta T$ between the arrival of two flashes of light from the particle
and concludes that $\Omega = 2\pi/\Delta T$ for the particle when the light
was emitted. This operational measure of $\Omega$ is independent of any gauge
choice made for $h^\act_{ab}$.

The components of the Killing vector $k^a$ in \Deqn{orbitk} are actually only
correct in a particular gauge for which both $\Lie_k g^0_{ab} = 0$ and
$\Lie_k h^\act_{ab} = 0$, individually. Under a gauge transformation the
coordinate description of $k^a$ changes by $\Or(\mu)$ with $\Delta k^a =
-\Lie_\xi k^a$. A choice for $\xi^a$ for which $\Lie_\xi k^a$ is not zero is
allowed but it would be very inconvenient and would result in a gauge for
which both $\Lie_k g^0_{ab} = \Or(h)$ and $\Lie_k h^\act_{ab} = \Or(h)$, even
though \Deqn{killing} would still hold. In principle, the geodesics of the
light rays could be computed from the particle out to the distant observer in
this inconvenient gauge and the orbital period could still be determined. But
in practice, this task would be horrendous.

In the convenient gauge, with $\Lie_k h^\act_{ab} = 0$, the calculation is
much easier. From the symmetry, it is clear that the change in Schwarzschild
coordinate time between the reception of two light flashes at the observer is
the same as the change in Schwarzschild coordinate time at the emission of
these flashes. Thus, $\Delta T$ measured operationally by a distant observer
is equal to the $\Delta T$ at the particle for one complete orbit, as long as
a gauge is used which respects the inherent symmetry of the example.

We next derive an expression for $\Omega$ in \Deqn{Omega} which is explicitly
gauge invariant for any transformation which respects the symmetry of the
example. This includes the possibility of a singular gauge transformation of
the type that would transform $h^\act_{ab}$ from the Regge-Wheeler gauge to
the Lorenz gauge.

We let the particle $\mu$ move along a geodesic of the perturbed
Schwarzschild geometry, $g_{ab} + h_{ab}$, where $h_{ab}$ is the regular
remainder $h^\R_{ab}$ for $\mu$ in a circular orbit in the equatorial plane.
The geodesic equation for the four-velocity of $\mu$ is
\begin{equation}
  \frac{\rd u_a}{\rd s} = \frac{1}{2} u^b u^c \frac{\partial}{\partial x^a}
       (g_{bc} + h_{bc})
\label{geod}
\end{equation}

The perturbation breaks the symmetries of the Schwarzschild geometry, and
there is no naturally defined energy or angular momentum for the particle.
However  we let $R(s)$ be the value of $r$ for the particle, and we define
specific components of $u_a$ by
\begin{equation}
  u_t = -E, \qquad u_\phi = J \qquad u^r = \dot R,\qquad\Text{and}\qquad
         u^\theta=0
\end{equation}
where $\dot{}$ denotes a derivative with respect to $s$. $E$ and $J$ are
reminiscent of the particle's energy and angular momentum per unit rest mass.

The components of the geodesic equation (\ref{geod}) are
\begin{eqnarray}
  \frac{\rd E}{\rd s}
     = -\frac{1}{2} u^a u^b \frac{\partial h_{ab}}{\partial t}
\label{dEds} \\
  \frac{\rd J}{\rd s}
     = \frac{1}{2} u^a u^b \frac{\partial h_{ab}}{\partial\phi}
\label{dJds} \\
  \frac{\rd}{\rd s}\Big(\frac{R\dot R}{R-2M} + u^a h_{ar}\Big) &=&
     \frac{1}{2} u^a u^b \frac{\partial}{\partial r} ( g_{ab} + h_{ab} ) .
\label{Rddot}
\end{eqnarray}
We are interested in the case when the orbit is nearly circular with $\dot R$
only resulting from the effects of energy and angular momentum loss. In this
case, $\dot E$ and $\dot J$ are $\Or(h)$, and we look for the additional
condition that $\dot R$ is also $\Or(h)$ to describe the slow inspiral of
$\mu$. All of the following equations in this section are assumed to be
correct through $\Or(h)$, unless otherwise noted.

The normalization of $u^a$ is a first integral of the geodesic equation, and
with the assumption that $\dot R = \Or(h)$ this is
\begin{equation}
 - u^a u^b (g_{ab} + h_{ab}) = 1 =  \frac{E^2}{1-2M/R} - \frac{J^2}{R^2}
      + {u}^a {u}^b h_{ab}.
\label{norm}
\end{equation}

While neither $\partial /\partial t$ nor $\partial /\partial \phi$ is a
Killing vector of $g_{ab}+h_{ab}$, the combination,
 $k^a = \partial /\partial t + \Omega \partial/\partial \phi$ is a Killing
vector in a preferred gauge, and ${u}^a$ is tangent to a trajectory of this
Killing vector, up to $\Or(h)$. Thus, at a circular orbit ${u}^a\partial_a
h_{bc} = \Or(h^2)$ in Schwarzschild coordinates.

A description of the quasi-circular orbits is obtained from \Deqn{Rddot} and
\Deqn{norm} by setting $\ddot R$ to zero. The results are
\begin{equation}
  E^2 = \frac{(R-2M)^2}{R(R-3M)} (1-{u}^a {u}^b h_{ab}
     - \frac{1}{2} R  {u}^a {u}^b \partial_r  h_{ab})
\label{E2}
\end{equation}
and
\begin{equation}
  J^2 = \frac{MR^2}{R-3M} (1-{u}^a {u}^b h_{ab})
   - \frac{R^3(R-2M)}{2(R-3M)}
        {u}^a {u}^b \partial_r  h_{ab}.
\end{equation}
Also the angular velocity, $\Omega$, of a circular orbit as measured at
infinity is
\begin{equation}
  \Omega^2 = (\rd \phi/\rd t)^2 = (u^\phi/u^t)^2 = M/R^3
    - \frac{R-3M}{2R^2} {u}^a {u}^b \partial_r h_{ab}.
\label{Omega}
\end{equation}
Finally,
\begin{equation}
  (E-\Omega J)^2 = (1-3M/R) (
        1- {u}^a {u}^b h_{ab}
        +R {u}^a {u}^b \partial_r h_{ab} /2).
\end{equation}

These equations give $E$, $J$, $\Omega$ and $E-\Omega J$ for a circular orbit
in terms of the radius of the orbit $R$ and the metric perturbation $h_{ab}$.
We can consider the effect on these expressions of a gauge transformation
which preserves the $\partial/\partial t + \Omega \partial/\partial \phi$
symmetry of the problem. The analysis uses descriptions of gauge
transformations found, for example, in references \cite{ReggeWheeler} and
\cite{Zerilli}. Here we present only the results.

The orbital frequency $\Omega$ and $E-\Omega J$ are both invariant under a
gauge transformation, while $E$ and $J$ are not. However, both $\rd E/\rd s$
and $\rd J/\rd s$ in \Deqn{dEds} and \Deqn{dJds} are gauge invariant. This
latter result might have been anticipated by using an operational definition
of energy and angular momentum loss as measured by a distant observer, and by
finding a relationship which joins the right hand sides of \Deqn{dEds} and
\Deqn{dJds} with the matching conditions at the particle for the differential
equations which describe the metric perturbation. This relationship is
straightforward but quite tedious to demonstrate directly.

The energy and the angular momentum measured by a distant observer are gauge
invariant. At zeroth order in the perturbation the energy is just $M$, the
mass of the black hole.  The $\el=0$ metric perturbation gives the $\Or(\mu)$
contribution to the energy and this is just $\mu E$, which is an $\Or(\mu)$
quantity independent of gauge but not relying upon the $\Or(\mu)$ terms in
\Deqn{E2}. The contribution of the gravitational self-force to the energy
measured at a large distance shows up only at second order in $\mu$; to
calculate this effect requires going to second order perturbation analysis.
Similar statements hold for the angular momentum measured at a large distance
and $J$.

For a circular orbit the radius, $R$, and both $E$ and $J$ all depend upon
the choice of gauge for $h_{ab}$.  However, $\Omega$ is defined in terms of a
measurement made at infinity, and $E-\Omega J$ is the contraction of $u_a$
with the Killing vector $\xi^a$; hence, these latter two quantities are
independent of the gauge, and this has been demonstrated explicitly allowing
for distribution valued gauge transformations.

The gauge invariance of $\Omega$ has an interesting twist. While $\Omega$ is
gauge invariant, the Schwarzschild radius of the orbit is not. A typical
gauge vector $\xi^a$ has a radial component which changes the coordinate
description $R$ of the orbit. This affects $\Omega$ through the $M/R^3$ term
in \Deqn{Omega}. This radial component of $\xi^a$ also changes the ${u}^a
{u}^b \partial_r h_{ab}$ in a manner that leaves the right hand side of
\Deqn{Omega} unchanged. Equation (\ref{Omega}) gives the same result whether
evaluated in a limit from outside the orbit or inside, in the event that
$h_{ab}$ is not differentiable at the orbit; this result follows from
analysis of the jump conditions on $h_{ab}$ at the orbit.

\section{Future prospects}
\label{conclusion}

Within the next year or two important applications of gravitational
self-force analyses will be viable.

For some time, it has been possible to calculate energy and angular momentum
loss by a small mass in an equatorial orbit about a Kerr black hole using the
Teukolsky \cite{Teukolsky73} formalism, which involves the Newman-Penrose
\cite{NewmanPenrose} scalars $\psi_0$ and $\psi_4$. For orbits off the
equatorial plane this is not good enough. For gravitational waveform
prediction, it is also necessary to calculate the dissipative change in the
third ``constant'' of the motion, the Carter constant $C$, due to
gravitational radiation. Energy and angular momentum loss may be determined
by finding the flux at a large distance in a gauge invariant manner. There is
no corresponding ``Carter constant flux.'' However, Lousto and Whiting
\cite{LoustoWhiting02,Lousto05} describe progress in determining metric
perturbations from $\psi_0$ or $\psi_4$. And Mino \cite{Mino03} has proposed
a method for determining $\mbox{d}C/\mbox{d}t$ which depends upon these
metric perturbations.

Self-force calculations in the Schwarzschild geometry are much easier, and
progress is likely to be rapid both in connecting results with post-Newtonian
analyses and in tracking the phase of gravitational radiation in an extreme
mass-ratio binary.

\subsection{Evolution of the phase during quasi-circular inspiral}

One application of self-force analysis is to track of the phase of the
gravitational wave from a small object while it orbits a large black hole
many times. For this task, Burko \cite{burko03} has emphasized the necessity
of using higher order perturbation theory to calculate properly the effects
of the conservative part of the self-force. Here, we follow his lead, and
estimate the number of orbits which can be tracked by use of analysis with
different levels of sophistication.

For definiteness, we assume that a small mass $\mu$ is undergoing slow,
quasi-stationary inspiral about a Schwarzschild black hole of mass $M$ and
that the orbit is relativistic so that $M$ gives the dynamical time scale. A
gauge-invariant $E$ of the orbiting particle is defined in terms of the mass
as measured at infinity,
\beq
  \mu E \equiv M_{\infty} - M .
\eeq
If we know $\Omega(E)$ and also $\mbox{d}E/\mbox{d}t$, then the assumption of
quasi-circular inspiral provides
\beq
  \frac{\mbox{d}\Omega}{\mbox{d}t}
  =  \frac{\mbox{d}\Omega}{\mbox{d}E} \frac{\mbox{d}E}{\mbox{d}t}.
\eeq
Let $\Omega_\oo$ be the orbital frequency at $t=0$. The phase of the orbit is
then
\bea
  \phi(t) &=& \int_0^t \Omega(E(t)) \,\mbox{d}t
\\ \nonumber
   &=& \int_0^t \left(\Omega_\oo + t \left[ \frac{\rd\Omega}{\rd E}
   \frac{\rd E}{\rd t}\right]  + \cdots
         \right)\rd t
\eea
after a Taylor expansion.

At the lowest level of approximation $\Omega$ and $E$ are given by the
geodesic equation in the Schwarzschild metric. The solution of the
first-order metric monopole perturbation problem, via Regge-Wheeler
\cite{ReggeWheeler} analysis, gives
\beq
 E \approx  E_\Text{\scriptsize 1st} \equiv -u_t \mbox{(circular orbit)}
\eeq
First order analysis, of the sort that was available in the 1970's, also
allows for the determination of $(dE/dt)_\Text{\scriptsize 1st}$. To obtain
new information \cite{burko03} regarding $E$, $\Omega$ and
$\mbox{d}E/\mbox{d}t$, requires second-order perturbation analysis, which
presupposes the solution of the first-order self-force problem. Second and
higher order analysis would provide
\beq
  \frac{\mbox{d}\Omega}{\mbox{d}t}
  =  \left[\frac{\mbox{d}\Omega}{\mbox{d}E}
    \frac{\mbox{d}E}{\mbox{d}t}\right]_\Text{\scriptsize 1st}
     \left[ 1 + \Delta_\Text{\scriptsize 2nd} + \cdots \right] ,
\eeq
where  $\Delta_\Text{\scriptsize 2nd} = \Or(\mu/\calR)$. With second or
higher order analysis, the phase is
\bea
  \phi
   &=& \int_0^t \left(\Omega_\oo + t \left[ \frac{d\Omega}{dE}
   \frac{dE}{dt}\right]_\Text{\scriptsize 1st}
          \left[1 + \Delta_\Text{\scriptsize 2nd} + \cdots \right] \right)dt
\nonumber\\
   &=& t \Omega_\oo + \frac{1}{2}t^2 \left[ \frac{d\Omega}{dE}
   \frac{dE}{dt}\right]_\Text{\scriptsize 1st}
          \left[1 + \Delta_\Text{\scriptsize 2nd} + \cdots \right]
\eea
after integration.

Consider the size of the contribution to the phase of the different terms of
\beq \fl
 \frac12 t^2 \left[ \frac{d\Omega}{dE} \frac{dE}{dt}
                         \right]_\Text{\scriptsize 1st}
             \left[1 + \Delta_\Text{\scriptsize 2nd} + \cdots \right]
  \approx \frac12 t^2 \left[\Or\left(\frac{\mu}{M^3}\right)
                      \right]_\Text{\scriptsize 1st}
             \left( 1 + \left[\Or\left(\frac{\mu}{M}\right)
                       \right]_\Text{\scriptsize 2nd}
                      + \cdots \right) .
\eeq
If radiation reaction is not included in the analysis, then none of this
term, of order $\frac12 t^2 \mu/M^3$, is included. This would lead to a phase
error of one full cycle after a time of order
 $t_\Text{\scriptsize dp} = M\sqrt{M/\mu}$, which is known as the
de-phasing timescale.

If only the first-order radiation reaction term is included, then the
$\frac12 t^2 \mu^2/M^4$ term is not included and leads to a phase error of
one full cycle after a time of order
 $t_\Text{\scriptsize rr} = M^2/\mu$. This is the radiation reaction
timescale.

If second-order radiation reaction is also included, then the $\cdots$ terms
of order $\frac12 t^2 \mu^3/M^5$ are not included and create a phase error of
one full cycle after a time of order
 $t_\Text{\scriptsize 2nd}  = (M^2/\mu) \sqrt{M/\mu})$.
This is the second-order timescale.

These same results are restated by noting that geodesic motion loses one
cycle of phase information after order $\sqrt{M/\mu}$ orbits.
 First order perturbation theory loses one
cycle of phase information after order $M/\mu$ orbits.
 And second order perturbation theory loses one
cycle of phase information after order $(M^2/\mu) \sqrt{M/\mu}$ orbits.

These estimates describe the difficulty involved in tracking the phase of an
orbit over an increasing number of orbits.

\subsection{Connection with post-Newtonian analyses}

An effort is now underway to find the effects of the gravitational self-force
on a number of parameters related to orbits in the Schwarzschild metric. The
first interesting results will be the orbital frequency $\Omega$ and the rate
of precession of the perihelion for a slightly eccentric orbit. Other
parameters which can be calculated with self-force analysis for circular
orbits are $E-\Omega J$ and a gauge invariant measure of the radius of the
orbit (see section \ref{gaugeinvariantquantities}).

First order perturbation theory coupled with self-force analysis will provide
the $\Or(\mu/M)$ effect on the innermost stable circular orbit (ISCO), as
well as the effect on the angular frequency of the ISCO. Currently, there is
no firm prediction as to whether the self-force moves the ISCO in or out.
Some recent scalar-field self-force results
\cite{RiveraMessaritakiWhitingDet04} show that the ISCO moves in and the
frequency of the ISCO increases; but there is no clear generalization of this
result to gravitation.

More interesting quantities will be available with second order perturbation
calculations, which now appear feasible. These include the energy, angular
momentum, and the rate of radiative loss of these quantities. Eventually,
second order gravitational wave-forms will be calculated.

One early goal of self-force analysis is to make contact with post-Newtonian
results. To do so requires that the quantities being calculated via
perturbation analysis match up precisely with those from post-Newtonian
analysis.

Post-Newtonian analyses are most reliable with slow speeds and weak
gravitational fields, and they easily accommodate comparable masses in a
binary. Perturbation analyses are most reliable with an extreme mass ratio,
but they accommodate fast speeds and strong fields. For, say, a $3M_\odot$
black hole orbiting a $20M_\odot$ black hole near its innermost orbit, the
mass ratio is not very extreme, the speeds are not very slow and the fields
are not very weak. Nevertheless, for this situation both post-Newtonian and
perturbation methods will be able to estimate properties of the system. A
comparison of these estimates will certainly highlight the strong and weak
aspects of each approach.

 \ack
My perspective on self-force effects in curved spacetime has developed over
the past few years from numerous discussions with my collaborators  Luz Maria
Diaz-Rivera, Dong-Hoon Kim, Eirini Messaritaki, Eric Poisson and Bernard
Whiting. In addition, the annual Capra meetings have provided a unique
atmosphere of collaborative enthusiasm, encouragement and stimulation. I am
deeply indebted to the organizers and participants of these fruitful
meetings.

This work was supported by the National Science Foundation under grant
PHY-0245024 with the University of Florida.

\appendix

\section{Perturbed Bianchi identity}
 \label{bianchi}

The Bianchi identity is
\beq
\nabla_c R_{dea}{}^b +\nabla_e R_{cda}{}^b +\nabla_d R_{eca}{}^b =0 .
\eeq
Contraction on $c$ and $b$
%%  yields
%%  \beq
%%  \nabla_b R_{dea}{}^b +\nabla_e R_{bda}{}^b + \nabla_d R_{eba}{}^b =0
%%  \eeq
%%  which
implies that
\beq
 \nabla_b R_{dea}{}^b =0
\label{vacuumbianchi}
\eeq
for a vacuum solution of the Einstein equations. This result is used often in
the derivations of identities involving $E_{ab}(h)$.

The definition of the operator $E_{ab}$ for a vacuum spacetime is
\beq \fl
2E_{ab}(h) = \nabla^2h_{ab} + \nabla_a\nabla_b h
          - 2\nabla_{(a}\nabla^c h_{b) c} + 2R_a{}^c{}_b{}^d h_{cd}
          + g_{ab} (\nabla^c\nabla^d h_{cd}-\nabla^2h) ,
\label{Eabbianchi}
\eeq
so that
\bea \fl
 2 \nabla^a E_{ab}(h) &=& \nabla^a\nabla^c\nabla_c h_{ab}
   + \nabla^a \nabla_a\nabla_b h
   - \nabla^a\nabla_{a}\nabla^c h_{b c}
   - \nabla^a\nabla_{b}\nabla^c h_{a c}
\nonumber\\  \fl && {}
   + 2(\nabla^a R_a{}^c{}_b{}^d) h_{cd}
   + 2 R^{ac}{}_b{}^d \nabla_a h_{cd}
   + \nabla_b\nabla^c\nabla^d h_{cd}
   - \nabla_b\nabla^c\nabla_c h .
%  \nonumber\\  &=&
%       \nabla^c\nabla^a\nabla_c h_{ab}
%                 + R^{ac}{}_c{}^d\nabla_d h_{ab}
%                 + R^{ac}{}_a{}^d\nabla_c h_{db}
%                 + R^{ac}{}_b{}^d\nabla_c h_{ad}
%  \nonumber\\ && {}
%     + R^a{}_{ba}{}^d \nabla_d h
%     - \nabla^a\nabla_{a}\nabla^c h_{b c}
%     - \nabla^a\nabla_{b}\nabla^c h_{a c}
%  \nonumber\\ && {}
%     + 2 R^{ac}{}_b{}^d \nabla_a h_{cd}
%     + \nabla_b\nabla^a\nabla^c h_{ac}
\nonumber\\ \fl  &=&
     \nabla^a\nabla^c\nabla_a h_{cb}
   - \nabla^a\nabla_{a}\nabla^c h_{b c}
%\nonumber\\ \fl  && {}
   - \nabla^a\nabla_{b}\nabla^c h_{a c}
   + \nabla_b\nabla^a\nabla^c h_{ac}
\nonumber\\ \fl  && {}
   + R^{ac}{}_b{}^d\nabla_c h_{ad}
   + 2 R^{ac}{}_b{}^d \nabla_a h_{cd}
%  \nonumber\\  &=&
%   \nabla^a (R^{c}{}_{ac}{}^d h_{db}) + \nabla^a (R^{c}{}_{ab}{}^d  h_{cd})
%             - R^a{}_{ba}{}^d \nabla^c h_{d c}
%   \nonumber\\ && {}
%     + R^{ac}{}_b{}^d\nabla_c h_{ad}
%     + 2 R^{ac}{}_b{}^d \nabla_a h_{cd}
  \nonumber\\ \fl  &=& 0 .
\label{pertBianchi}
\eea
The second equality follows after use of the Ricci identity to interchange
the order of derivatives on the first, second and last terms as well as
repeated uses of $R_{ab}=0$ and \Deqn{vacuumbianchi} for vacuum spacetimes.
The final result follows after use of the Ricci identity on the first two
terms and on the second two terms of the second equality, and the application
of symmetries of the Riemann tensor on the remainder.

If $h_{ab}$ is not $C^3$ then $\nabla^a E_{ab}(h) = 0$ in a distributional
sense. To show this, choose an arbitrary, smooth test vector field
$\lambda^a$ with compact support. Consider the integral of $\lambda^b
\nabla^a E_{ab}(h)$ over a sufficiently large region. Integrate by parts once
and discard the surface term. Next use \Deqn{EabGreen} and discard the
surface terms to obtain an integral of $h^{ab} E_{ab}(\nabla\lambda)$. This
integral is zero from \Deqn{gaugeT}. These steps also provide an alternative
derivation of \Deqn{pertBianchi} in the event that $h_{ab}$ is, in fact,
$C^3$.

%  \noindent First RHS:
%  \\ \indent $\nabla^a R_a{}^c{}_b{}^d = 0$ for vacuum spacetimes.
%  \\ \indent Ricci identity on the first term
%  \\ \indent Ricci identity on second and last terms
%  \\ \indent relabel indices on the penultimate term.
%
%  \noindent Second RHS:
%  \\ \indent $R_{ab}= 0$ for vacuum spacetimes, three times.
%  \\ \indent Relabel indices $a\leftrightarrow c$ in the first term.
%
%  \noindent Third RHS:
%  \\ \indent Ricci identity on the first and second terms and on the third
%  and fourth terms.
%
%  \noindent Fourth RHS:
%  \\ \indent $R_{ab}= 0$ twice and $\nabla^a R^c{}_{ac}{}^d = 0$ for
%  vacuum spacetimes.
%  \\ \indent Symmetries of the Riemann tensor for the remainder
%  of the terms.

\section{Formal $\mathbf{n}$th order perturbation analysis}
\label{nthorder}

In general perturbation analysis, let  the $g_{ab}$ of \Deqn{Eab} be an exact
solution to the vacuum Einstein equations, $g^0_{ab}$, and iteratively define
\begin{equation}
  g^{(n)}_{ab} = g^{(n-1)}_{ab} + h^{(n)}_{ab}
\end{equation}
where
\begin{equation}
  h^{(n)}_{ab} = \Or(\mu^n).
\end{equation}
Assume that we are given $g^{(n-1)}_{ab}$ and $T^{(n)}_{ab} = \Or(\mu)$, with
\begin{equation}
  G^{(n-1)}_{ab} - 8\pi T^{(n)}_{ab} = \Or(\mu^{n}).
\label{Gabn}
\end{equation}
If $h^{(n)}_{ab}$ is a solution of \Deqn{Gabn} from
\begin{equation}
  E_{ab}(h^{(n)}) = G^{(n-1)}_{ab} - 8\pi T^{(n)}_{ab} + \Or(\mu^{n+1}) ,
\label{Eabn}
\end{equation}
then it follows from the definition of the operator $E_{ab}(h)$ in
\Deqn{Eabdef} that
\begin{equation}
  G^{(n)}_{ab} - 8\pi T^{(n)}_{ab} = \Or(\mu^{n+1}),
\end{equation}
and $h^{(n)}_{ab}$ is an $\Or(\mu^n)$ improvement to the approximate solution
to the Einstein equations.

The Bianchi identity implies that
\begin{equation}
  \nabla^a E_{ab}(h) = 0
\label{divEn}
\end{equation}
 for any symmetric $C^3$ tensor field $h_{ab}$, as shown in \ref{bianchi}.
It is also shown that if $h_{ab}$ is not $C^3$ then \Deqn{divEn} holds in a
distributional sense.
 Thus an integrability condition of \Deqn{Eabn} is that
\begin{equation}
  \nabla^a \big(G^{(n-1)}_{ab} - 8\pi T^{(n)}_{ab}\big) = \Or(\mu^{n+1}).
\label{divG1}
\end{equation}
Note, however, that
\begin{eqnarray}
\fl  \nabla^a \big(G^{(n-1)}_{ab} - 8\pi T^{(n)}_{ab}\big) & = &
  \nabla^a_{(n-1)} \big(G^{(n-1)}_{ab} - 8\pi T^{(n)}_{ab}\big)
\nonumber\\ &&
  {} + \Gamma^a_{ac}\big(G^{(n-1) c}_b - 8\pi T^{(n)c}_b\big)
% \nonumber\\ &&
  {} - \Gamma^c_{ab}\big(G^{(n-1) a}_c - 8\pi T^{(n)a}_c \big),
\label{divG}
\end{eqnarray}
where $\nabla^a_{(n-1)}$ is the derivative operator of $g^{(n-1)}_{ab}$, and
$\Gamma^a_{bc}$ is the connection relating the derivative operators
$\nabla^a$ and $\nabla_{(n-1)}^a$.
 The Bianchi identity implies that
\begin{equation}
  \nabla^a_{(n-1)}G^{(n-1)}_{ab} = 0,
\end{equation}
and the terms in \Deqn{divG} involving $\Gamma^a_{bc}$ are order $\mu^{n+1}$
because of \Deqn{Gabn} and the fact that $\Gamma^a_{bc} = \Or(\mu)$. Thus,
the approximate vanishing of the right hand side of \Deqn{divG} is the
integrability condition for \Deqn{Eabn},
\begin{equation}
  \nabla^a_{(n-1)}T^{(n)}_{ab} = \Or(\mu^{n+1}).
\label{divTn}
\end{equation}

In other words, before \Deqn{Eabn} can be solved for $h^{(n)}_{ab}$, it is
necessary that the perturbing stress tensor be adjusted to be conserved with
the metric $g^{(n-1)}_{ab}$ and to satisfy \Deqn{divTn}.

\section{$\nabla_b T^{ab} = 0$ implies the
  geodesic equation for a point mass} \label{geodesic}

We follow an example in reference \cite{problembook}. In \Deqn{Tab},
$\delta^4(x^a-X^a(s))/\sqrt{-g}$ is a scalar field, and the factor $u^b$ may
be defined as a vector field by extension, in any smooth manner, away from
the world line. Then,
\bea
  (g^c{}_a+u^cu_a) \nabla_b T^{ab} &=&
    \mu (g^c{}_a+u^cu_a)
    \int_{-\infty}^\infty
    \left[ \frac{ (\nabla_b u^a) u^b}{\sqrt{-g}} \delta^4(x^a-X^a(s)) \right.
\nonumber\\ && \left.
    {}+ u^a \nabla_b \left(\frac{u^b}{\sqrt{-g}}
          \delta^4(x^a-X^a(s))\right) \right] \,\mbox{d}s
\nonumber\\ &=&
     \mu \int_{-\infty}^\infty
        \frac{ (\nabla_b u^a) u^b}{\sqrt{-g}} \delta^4(x^a-X^a(s))
                 \,\mbox{d}s
 \eea
where the second equality follows from properties of the projection operator
$g^c{}_a+u^cu_a$. If $\nabla_b T^{ab} = 0$, then it necessarily follows that
the coefficient of the delta function must be zero for all proper times. A
consequence is that  $u^b \nabla_b u^a = 0$, the geodesic equation.

A more formal proof of this result is in Poisson's review of the self-force
\cite{poisson:03}, p 89.
% Poisson's Living Review p89, eqns.~(4.3.2), (12.1.3) and (18.1.3).

\section{Gauge invariance of $E_{ab}(h)$}
 \label{gaugeinvariance}

For a background geometry which is a vacuum solution of the Einstein
equations, an infinitesimal gauge transformation, $x^a_{\Text{\scriptsize
new}} = x^a + \xi^a$, with $\xi^a = \Or(\mu)$ changes the metric
perturbation, $h^{\Text{\scriptsize new}}_{ab} = h_{ab} - 2\nabla_{(a}
\xi_{b)} + \Or(\mu^2)$.  But the operator $E_{ab}(h)$ is invariant under such
a coordinate transformation,
\begin{equation}
  E_{ab}(\nabla \xi) = 0.
\label{gaugeT}
\end{equation}
This result follows immediately from the fact that the change in the
perturbation of the Einstein tensor $E_{ab}$ under a gauge transformation is
the Lie derivative of the background Einstein tensor $\Lie_\xi G_{ab}$. For a
vacuum background spacetime, this is zero.

Equation (\ref{gaugeT}) also follows from direct substitution into
\beq \fl
2E_{ab}(h) = \nabla^2h_{ab} + \nabla_a\nabla_b h
          - 2\nabla_{(a}\nabla^c h_{b) c} + 2R_a{}^c{}_b{}^d h_{cd}
          + g_{ab} (\nabla^c\nabla^d h_{cd}-\nabla^2h)
\label{Eabgauge}
\eeq
with $ h_{ab} = 2\nabla_{(a}\xi_{b)}$.
 It is easiest to consider the factor of $g_{ab}$ separately,
\bea
 \Text{factor of } g_{ab} &=&
    \nabla^c\nabla^d \nabla_c\xi_d
    + \nabla^c\nabla^d \nabla_d\xi_c
  - 2 \nabla^a\nabla_a\nabla^b\xi_b
%  \nonumber\\ &=&
%      \nabla^c\nabla^d \nabla_c\xi_d
%      + \nabla^d\nabla^c \nabla_d\xi_c + R^{cd}{}_d{}^e  \nabla_e\xi_c
%                                       + R^{cd}{}_c{}^e  \nabla_d\xi_e
%    - 2 \nabla^a\nabla_a\nabla^b\xi_b
\nonumber\\ &=&
    2 \nabla^c\nabla^d \nabla_c\xi_d
  - 2 \nabla^a\nabla_a\nabla^b\xi_b
%  \nonumber\\ &=&
%      2 \nabla^c\nabla^d \nabla_c\xi_d
%    - 2 \nabla^a\nabla^b\nabla_a\xi_b - 2\nabla^a[R_a{}^b{}_b{}^e \xi_e]
%  \nonumber\\ &=&
%      2 \nabla^c\nabla^d \nabla_c\xi_d
%    - 2 \nabla^a\nabla^b\nabla_a\xi_b
\nonumber\\ &=& 0 .
\eea
The second equality follows after use of the Ricci identity on the first two
indices of the second term, use of $R_{ab}=0$ for a vacuum spacetime and a
relabeling of the indices. The final result follows after use of the Ricci
identity on the second term of the second equality and use of $R_{ab}=0$ for
a vacuum spacetime.
%  \noindent First RHS: Ricci identity on second term\\
%  Second RHS: $R^{de}=0$, relabel second term\\
%  Third RHS: Ricci identity on second term\\
%  Fourth RHS: $R^{de}=0$\\
%  Fifth RHS: relabel indices.
%  %
With $ h_{ab} = 2\nabla_{(a}\xi_{b)}$, the remainder of $E_{ab}(2\nabla\xi)$
is
\bea  \fl
 \Text{remainder} &=&
      \nabla^c\nabla_c\nabla_a\xi_b
    + \nabla^c\nabla_c\nabla_b\xi_a
%\nonumber\\  \fl && {}
    + 2 \nabla_a\nabla_b \nabla^c\xi_c
\nonumber\\  \fl && {}
    - \nabla_{a}\nabla^c \nabla_b\xi_c
    - \nabla_{a}\nabla^c \nabla_c\xi_b
%\nonumber\\  \fl && {}
    - \nabla_{b}\nabla^c \nabla_a\xi_c
    - \nabla_{b}\nabla^c \nabla_c\xi_a
\nonumber\\  \fl && {}
   + 2R_a{}^c{}_b{}^d \nabla_c\xi_d   + 2R_a{}^c{}_b{}^d \nabla_d\xi_c .
\eea
The analysis of this expression is lengthy but not difficult. It begins with
using the Ricci identity
 upon the second and third indices of the first, second, fourth and sixth
terms and
 upon the first and second indices of the fifth and seventh terms.
The resulting terms with three derivatives may be paired up in a way to use
the Ricci identity again and to reduce the entire expression to one involving
only single derivatives. This also requires application of
\Deqn{vacuumbianchi}. That the entire expression is zero, then follows from
the symmetries of the Riemann tensor.

\section{Green's theorem for $E_{ab}$ }
 \label{fluxintegral}

\newcommand{\K}{{k}}

The operator $E_{ab}(h)$ in \Deqn{Eab}, with an arbitrary tensor $k^{ab}$,
satisfies the identity
\begin{equation}
  2 k^{ab} E_{ab}(h) =
      \nabla_c F^c(k,h) - \big<k^{ab}, h_{ab}\big> ,
\label{divF}
\end{equation}
where
\begin{eqnarray}  \fl
  F^c(k,h) &\equiv& k^{ab} \nabla^c h_{ab}
  - \frac{1}{2} k \nabla^c h
%\nonumber\\  \fl &&
       - 2 (k^{cb} - \frac{1}{2}g^{cb}k)
                       \nabla^a(h_{ab} - \frac{1}{2}g_{ab}h)
\label{Fc}
\end{eqnarray}
and
\begin{eqnarray}  \fl
  \big<k^{ab}, h_{ab}\big> & \equiv &
        \nabla^ck^{ab} \nabla_c h_{ab}
      - \frac{1}{2}\nabla^ck \nabla_c h
\nonumber\\ \fl  &&
      {} - 2 \nabla_a(k^{ac}-\frac{1}{2}g^{ac}k)
              \nabla^b(h_{bc}-\frac{1}{2}g_{bc}h)
% \nonumber \\ &&
      {}         - 2 k^{ab} {{{{R_a}^c}_b}^d} h_{cd} .
\label{hbarh}
\end{eqnarray}
Note that the ``inner product,''
 $\big<k^{ab}, h_{ab}\big> = \big<h^{ab}, k_{ab}\big>$
is symmetric under the interchange of $h^{ab}$ and $k_{ab}$. It follows that
\begin{equation}
   k^{ab} E_{ab}(h) - h^{ab} E_{ab}(k)
    =\frac{1}{2}  \nabla_c \left[ F^c(k,h) - F^c(h,k) \right] .
\label{EabGreen}
\end{equation}
Which is a tensor version of Green's theorem for the differential operator
$E_{ab}(h)$.

The derivation of equation (\ref{divF}) is straightforward. Contract
\Deqn{Eab} with an arbitrary symmetric tensor $k^{ab}$, and move $k^{ab}$
inside $\nabla_a$ in each term by ``differentiating by parts.'' the
divergence terms determine $F^c(k,h)$.

\section{Singular gauge transformations}
 \label{singulargauge}
Let $\xi^a$ be a, possibly distribution valued, vector field. And let $h_{ab}
= - 2 \nabla_{(a}\xi_{b)}$, as for a gauge transformation. Also, let $k_{ab}$
be a smooth ``test'' tensor with compact support. Then
\bea
  \int k^{ab}E_{ab}(h) \sqrt{-g} \, \mbox{d}^4x &=&
%  \int \nabla_c [F^c(k,h)-F^c(h,k)]\sqrt{-g} \,\mbox{d}^4x
%\nonumber\\ && {} +
       \int h^{ab}E_{ab}(k)\sqrt{-g} \, \mbox{d}^4x
\nonumber\\ &=&
       - 2 \int (\nabla^a \xi^b) E_{ab}(k)\sqrt{-g} \, \mbox{d}^4x ,
\eea
from \Deqn{EabGreen}, after dropping the divergence term. An integration by
parts and application of the perturbed Bianchi identity (\ref{pertBianchi})
yields
\bea
  \int k^{ab}E_{ab}(h) \sqrt{-g} \,\mbox{d}^4x &=&
        2 \int \xi^b \nabla^a \left[ E_{ab}(k) \right]
            \sqrt{-g} \,\mbox{d}^4x
\nonumber\\ &=& 0 .
\eea
Thus, we demonstrate that given a solution to the inhomogeneous, perturbed
Einstein equations \Deqn{EabTab}, even a distributional gauge transformation
leaves a distributional valued metric perturbation that continues to satisfy
the perturbed Einstein equations in this distributional sense.

\section{Black hole moving through an external background geometry}
 \label{SloMo}

When a small Schwarzschild black hole of mass $m$ moves through a background
spacetime, the hole's metric is perturbed by quadrupole tidal forces arising
from ${}_2H_{ab}$ in \Deqn{H2} or \Deqn{H2new}, and the actual metric near
the black hole, including the quadrupole perturbation, is
\begin{equation}
  g^\act_{ab} = g_{ab}^{\Text{\scriptsize Schw}} + {}_2h_{ab}
   + \Or( r^4/\calR^4),
\label{gabpertAPP}
\end{equation}
where the quadrupole metric perturbation ${}_2h_{ab}$ is a solution of
\begin{equation}
 E^{\Text{\scriptsize Schw}}_{ab}({}_2h) = 0 .
\label{hlabAPP}
\end{equation}
Here $E^{\Text{\scriptsize Schw}}_{ab}$ is the Schwarzschild geometry version
of the operator given in \Deqn{Eab}. The appropriate boundary conditions for
\Deqn{hlabAPP} are that the perturbation be well behaved on the event horizon
and that ${}_2h_{ab} \rightarrow {}_2H_{ab}$ in the buffer region, where $\mu
\ll r \ll \calR$.

Our analyses of the boundary conditions and solutions of equation
(\ref{hlabAPP}) for slow motion are very strongly influenced by Poisson's
recent analysis \cite{Poisson04a,Poisson04b,Poisson05} of the same situation.
In this appendix we describe ${}_2h_{ab}$ up to a remainder of
$\Or(r^4/\calR^4)$.

The appropriate boundary conditions at the future event horizon are that
$h_{ab}$ be finite and well behaved in a coordinate system which is well
behaved itself.  The Eddington-Finkelstein ingoing coordinates are
satisfactory, and
\beq
  V = t + r_\ast \quad \Text{and} \quad R=r,
\label{VRtrAPP}
\eeq
where $r_\ast = r+2m\ln(r/2m-1)$; the angles $\theta$ and $\phi$ remain
unchanged.

The odd and even parity parts of the metric perturbation are governed by the
Regge-Wheeler \cite{ReggeWheeler} and  Zerilli \cite{Zerilli} equations,
respectively.
 For our purposes, we change these equations from the Schwarzschild $t$,$r$ to
the Eddington-Finkelstein $V,R$. For example, the Regge-Wheeler equation for
$W(V,R)$ becomes
\beq
  2  \frac{\partial^2 W}{\partial V \partial R}
 + \left(1-\frac{2m}{R}\right) \frac{\partial^2 W}{\partial R^2}
 + \frac{2m}{R^2} \frac{\partial W}{\partial R}
 - 6(R - m) \frac{W}{R^3} = 0 ,
\label{RWapp}
\eeq
where we have assumed that the angular dependence of $W$ corresponds to a
linear combination of $\el=2$ spherical harmonics. We next assume that $W$ is
slowly changing in $V$ and accordingly let
\beq
  W(V,R) = \calB W_0(R) + \calB' W_1(R) + \ldots
  \quad\Text{where} \quad \calB' = d\calB(V)/dV .
\label{WVR}
\eeq
$\calB$ is a function of $V$, and of the angles, that is related ultimately
to the time-dependent external quadrupole moment of the geometry through
which the black hole is moving. Thus $\calB = \Or(\calR^{-2})$ and  $\calB' =
\Or(\calR^{-3})$, in keeping with the requirement of slow time dependence.

With the form \Deqn{WVR} substituted into \Deqn{RWapp}, we separate the terms
of $\Or(\calR^{-2})$ from those of $\Or(\calR^{-3})$ to obtain two equations.
One is an ordinary, homogeneous differential equation for $W_0(R)$, and the
second is for for $W_1(R)$ with a source term from the $\partial^2 (\calB
W_0)/\partial V \partial R$ term. An analytic solution of the first equation
is $W_0(R)=R^3$. The solution of the second for $W_1(R)$ is also analytic but
more complicated, and constants of integration may be chosen so that $W_1$ is
well behaved at the horizon. This procedure thus provides a general solution
for $W(V,R)$, up to a remainder of $\Or(\calR^{-4})$.  The even parity
Zerilli equation may be solved in a similar manner or by using the simple
relationship between solutions of the Regge-Wheeler and Zerilli equations
\cite{ChandraDet75}.

From the solutions of the Regge-Wheeler and Zerilli equations, the actual
metric perturbations are determined by taking derivatives of the master
variables. This results in a metric perturbation, in the Regge-Wheeler gauge,
whose non-zero Schwarzschild components, as functions of $V$ and $r$, are
%% \textbf{\\Elementary slow motion solution,
%%           not in the true-harmonic gauge:
%% based on \texttt{SlowMoSimpExpFeb7.mws}, with $b1=-49/10$}
\bea % A =
\fl
  h_{tt} &=&   - (r - 2m)^2 [ \calE^{(2)}  - 2m\ln(r/2m) \,\calE'^{(2)}]
\nonumber\\ \fl &&
   {} + \frac{1}{3r^2}
      (3 r^5 - 12 r^4 m  + 36 m^3r^2 - 16 m^4r - 8 m^5) \calE'^{(2)}
\label{Aapp}
\eea
\bea % -D =
  \fl  h_{tr} = -\frac{r(2r^3 - 3mr^2 - 6m^2r + 6m^3)}{3(r-2m)}\calE'^{(2)}
\eea
\bea % E =
  \fl \frac{1}{2}\h^\trc &=& -(r^2 - 2m^2) [\calE^{(2)} -2m \ln(r/2m) \,\calE'^{(2)} ]
\nonumber\\ \fl && {}
    + \frac{1}{3r}(3 r^4 - 18 m^2r^2 - 12 m^3r + 8 m^4) \calE'^{(2)}
\eea
\bea  % K =
  \fl  h_{rr} &=& - r^2 [ \calE^{(2)}  - 2m \ln(r/2m) \,\calE'^{(2)}]
\nonumber\\ \fl && {}
   + \frac{1}{3}
       \frac{(3r^5 - 12r^4m + 36 m^3r^2 - 16m^4r - 8m^5)}{(r - 2m)^2}
     \calE'^{(2)}
\eea
\bea  % - C =
  \fl h^\od_{tA} &=& \frac{1}{3}r(r - 2m)
                  [\calB^{(2)}_A - 2m\ln(r/2m) \,\calB'^{(2)}_A]
\nonumber\\ \fl && {}
     - \frac{1}{9r^2}
   (3 r^5 - 6 r^4m - 12 r^3m^2 + 12 r^2m^3 + 8 rm^4 + 8 m^5) \calB'^{(2)}_A
\eea
\bea %  J =
  \fl h^\od_{rA} = \frac{r^4}{12(r - 2m)} \calB'^{(2)}_A .
\label{Japp}
\eea
This metric perturbation was first derived by Poisson \cite{Poisson05} in a
different gauge.

In these expressions $\calB^{(2)}$ and  $\calE^{(2)}$ are $V$-dependent
linear combinations of the $\el=2$ spherical harmonic functions
$Y_{2,m}(\theta,\phi)$. The $V$ and $R$ coordinate components are all well
behaved on the future event horizon. The $V$-dependence of $\calE^{(2)}$ and
$\calB^{(2)}$ shows that the metric perturbation propagates toward the black
hole from a great distance as expected.

To make contact with the actual, external geometry it is useful to expand the
expressions given in \Deqn{Aapp}-\Deqn{Japp} for $r$ in the buffer region,
where $m \ll r \ll \calR$, and we take advantage of the fact that
$\calB'^{(2)}$ and  $\calE'^{(2)}$ are $\Or(\calR^{-3})$. Thus, for $r_\ast
\ll \calR$ a Taylor series about $V=t$ provides
\bea
    \calB(V) &=& \calB(t+r_\ast)
\nonumber\\
    &=& \calB(t) + r_\ast \frac{d\calB(t)}{dt} + \Or(\calR^{-4}) .
\eea
For $m \ll r \ll \calR$, the even parity part of the metric perturbation in
Schwarzschild coordinates is
\bea \fl
   h^\ev_{ab} \dx^a \dx^b &=& - \calE^{(2)}
      \left[(r - 2m)^2 \dt^2 +r^2 \dr^2 + (r^2-2m^2)\sigma_{AB} \dx^A
      \dx^B\right]
\nonumber\\ \fl &&
  {} + \frac{16m^6}{15r^4} \dot\calE^{(2)}
      \left[  2(r+m) \dt^2
            + 2(r+5m) \dr^2
            + (2r+5m)\sigma_{AB} \dx^A\dx^B\right]
\nonumber\\ \fl && % h_{tr} =
 {} - 2 \frac{r(2r^3 - 3mr^2 - 6m^2r + 6m^3)}{3(r-2m)}
            \dot\calE^{(2)} \dt\dr
  + \Or(m^8\dot\calE^{(2)}/r^5).
\label{hevapp}
\eea
and the odd parity part is
\bea
\fl  h^\od_{ab} \dx^a \dx^b &=&
       2 \left[ \frac{r}{3}(r-2m) \calB^{(2)}_A
              + \frac{16m^6}{45r^4} (3r+4m) \dot\calB^{(2)}_A \right]
     \dt \dx^A
\nonumber\\ \fl &&  %  h^\od_{rA} =
 {} + 2 \frac{r^4}{12(r - 2m)} \dot\calB^{(2)}_A \dr \dx^A
  + \Or(m^8\dot\calB^{(2)}/r^5).
\label{hodapp}
\eea
In this form $\calE^{(2)}$ and $\calB^{(2)}$ are considered functions of $t$
and $\dot\calE^{(2)}$ denotes the $t$ derivative of $\calE^{(2)}$.

\section*{References}

%  \bibliographystyle{prsty}
%  \bibliography{BBH}

\end{document}